\documentclass[twocolumn]{aastex6}
\usepackage{graphics}
\usepackage{grffile}
\usepackage{rotating}
\usepackage{amsmath}

\usepackage{natbib}
\usepackage{color}
\usepackage{times}
\usepackage[normalem]{ulem}             
\usepackage{cancel}
\usepackage{verbatim}				
\usepackage{mathrsfs}
\usepackage[mathscr]{euscript}

\newcommand{\DM}{\rm DM}
\newcommand{\SM}{\rm SM}
\newcommand{\EM}{\rm EM}

\newcommand{\cnsq}{\rm C_n^2}
\newcommand{\dnud}{\Delta\nu_d}                 
\newcommand{\taud}{\tau_d}                      
\newcommand{\taudbar}{\overline{\tau}_d}        
\newcommand{\ds}{D_s}

\newcommand{\be}{\begin{eqnarray}}
\newcommand{\ee}{\end{eqnarray}}

\tighten

\begin{document}

\newcommand{\numbering}{\nonumber}

\newcommand{\rF}{r_{\rm  F}}
\newcommand{\phiF}{\phi_{\rm  F}}

\newcommand{\linner}{l_{\rm  linner}}
\newcommand{\louter}{l_{\rm louter}}

\newcommand{\qinner}{q_{\rm inner}}
\newcommand{\qouter}{q_{\rm outer}}

\newcommand{\gr}{g_{\rm r}}
\newcommand{\gd}{g_{\rm d}}

\newcommand{\md}{m_{\rm d}}
\newcommand{\mr}{m_{\rm r}}
\newcommand{\mg}{m_{\rm g}}

\newcommand{\mdb}{m_{\rm d, B}}
\newcommand{\mdtheta}{m_{\rm d, \theta}}
\newcommand{\mdbtheta}{m_{\rm d, B\theta}}
\newcommand{\Deff}{D_{\rm eff}}

\newcommand{\SN}{\rm S/N}
\newcommand{\Sp}{S^{\,\prime}}
\newcommand{\St}{{S_{\rm t}}}			
\newcommand{\Ntrials}{N_{\rm trials}}
\newcommand{\Ntrialsmax}{N_{\rm trials}(\rm max)}
\newcommand{\Ndiss}{N_{\rm DISS}}

\newcommand{\Nd}{N_{\rm d}}
\newcommand{\tp}{t^{\,\prime}}
\newcommand{\NF}{N_{\rm F}}
\newcommand{\TF}{T_{\rm F}}
\newcommand{\Rdf}{R_{\rm df}}

\newcommand{\Ts}{T_{\rm s}}
\newcommand{\Nbeams}{N_{\rm \Omega}}
\newcommand{\Ms}{M_{\rm s}}
\newcommand{\MF}{M_{\rm F}}
\newcommand{\pdfx}{f_{\rm x}}
\newcommand{\pdfy}{f_{\rm y}}


\newcommand{\DMfrb}{\DM_{\rm frb}}
\newcommand{\DMg}{\DM_{\rm g}}
\newcommand{\DMh}{\DM_{\rm h}}
\newcommand{\dDMc}{\delta\DM_{\rm c}}
\newcommand{\DMigm}{\DM_{\rm igm}}
\newcommand{\DMscatt}{\DM_{\rm scatt}}

\renewcommand{\dnud}{\Delta\nu}

\renewcommand{\taud}{\tau}
\newcommand{\taufrb}{\tau_{\rm frb}}
\newcommand{\taudh}{\tau_{\rm h}}
\newcommand{\taudg}{\tau_{\rm g}}
\newcommand{\taudigm}{\tau_{\rm igm}}

\newcommand{\etah}{\eta_{\rm h}}
\newcommand{\etac}{\eta_{\rm c}}
\newcommand{\etaigm}{\eta_{\rm igm}}
\newcommand{\etag}{\eta_{\rm g}}

\newcommand{\Lh}{L_{\rm h}}
\newcommand{\Lc}{L_{\rm c}}
\newcommand{\Ligm}{L_{\rm igm}}
\newcommand{\Lg}{L_{\rm g}}

\newcommand{\Fh}{F_{\rm h}}
\newcommand{\Fc}{F_{\rm c}}
\newcommand{\Figm}{F_{\rm igm}}
\newcommand{\Fg}{F_{\rm g}}

\newcommand{\DMism}{\DM_{\rm ism}}
\newcommand{\DMpsr}{\DM_{\rm psr}}
\newcommand{\DMhalo}{\DM_{\rm halo}}

\newcommand{\taudism}{\tau_{\rm ism}}
\newcommand{\taudpsr}{\tau_{\rm psr}}
\newcommand{\taudpsrbar}{\overline{\tau}_{\rm psr}}
\newcommand{\taudc}{\tau_{\rm c}}
\newcommand{\dtaudc}{\delta\tau_{\rm c}}
\newcommand{\sigmapsr}{\sigma_{\rm psr}}
\newcommand{\xic}{\xi_{\rm c}}

\newcommand{\Ftilde}{\widetilde F}
\newcommand{\Ftism}{\widetilde F_{\rm ism}}
\newcommand{\Ftc}{\widetilde F_{\rm c}}
\newcommand{\rc}{r_{\rm c}}
\newcommand{\wc}{w_{\rm c}}
\renewcommand{\sc}{s_{\rm c}}

\newcommand{\taudhat}{\widehat\taud}
\renewcommand{\taudbar}{\overline{\taud}}
\renewcommand{\linner}{l_{\rm i}}
\renewcommand{\louter}{l_{\rm o}}
\newcommand{\dnc}{d_0}
\newcommand{\dwc}{d_{\rm wc}}
\newcommand{\nezero}{n_{e_0}}
\newcommand{\CSM}{{C_{\SM}}}
\newcommand{\ff}{f_{\rm f}}

\newcommand{\etaism}{\eta_{\rm ism}}
\newcommand{\Fth}{\widetilde F_{\rm h}}
\newcommand{\Ftigm}{\widetilde F_{\rm igm}}
\newcommand{\Ftg}{\widetilde F_{\rm g}}

\newcommand{\deff}{d_{\rm eff}}
\newcommand{\DMxgal}{\DM_{\rm xg}}
\newcommand{\DMgal}{\DM_{\rm g}}

\newcommand{\tauism}{\tau_{\rm ism}}
\newcommand{\tauh}{\tau_{\rm h}}
\newcommand{\tauc}{\tau_{\rm c}}

\newcommand{\tauxg}{\tau_{\rm xg}}
\newcommand{\taug}{\tau_{\rm g}}

\newcommand{\zg}{z_{\rm g}}

\renewcommand{\ds}{d_{\rm s}}

\newcommand{\nec}{n_{\rm e_c}}
\newcommand{\nesc}{n_{\rm e_{sc}}}
\newcommand{\DMc}{{\DM_{\rm c}}}
\newcommand{\SMc}{{\SM_{\rm c}}}
\newcommand{\EMc}{{\EM}_{\rm c}}

\title{Radio Wave Propagation and the Provenance of Fast Radio Bursts}
\author{
J.~M.~Cordes\altaffilmark{1,2},
R.~S.~Wharton\altaffilmark{1},
L.~G.~Spitler\altaffilmark{2},
S. Chatterjee\altaffilmark{1},
I. Wasserman\altaffilmark{1}
}
\altaffiltext{1}{Dept. of Astronomy and Cornell Center for Astrophysics and Planetary Science, Cornell Univ., Ithaca, NY 14853, USA}
\altaffiltext{2}{Max-Planck-Institut f\"ur Radioastronomie, Auf dem H\"ugel 69, 53121 Bonn, Germany}


\begin{abstract}
We analyze plasma dispersion and scattering of fast radio bursts (FRBs) to identify the dominant locations of free electrons along their  lines of sight and thus constrain the distances of the burst sources themselves.  We establish the  average $\tau$-DM relation for Galactic pulsars and use it as a benchmark for discussing FRB scattering.  Though  scattering times $\tau$ for   FRBs  are  large in the majority of the 17 events we analyze,  they are systematically smaller than those of Galactic pulsars that have similar dispersion measures (DMs).     The lack of any correlation between $\tau$ and $\DM$ for FRBs suggests that the intergalactic medium (IGM) cannot account for both $\tau$ and $\DM$. We therefore consider mixed models involving  the IGM and host galaxies.   If the IGM contributes significantly to DM  while  host galaxies dominate $\tau$, the scattering deficit with respect to the mean Galactic trend can be explained with a $\tau$-DM relation in the host that matches  that for the Milky Way.  However,  it is possible that hosts dominate both $\tau$ and $\DM$, in which case the observed scattering deficits require  free  electrons in the host to be less turbulent than in the Galaxy, such as if they are in hot rather than warm ionized regions.
Our results imply that  distances or redshifts of FRB sources can be significantly overestimated if they are  based on the assumption that the extragalactic portion of DM is dominated by the IGM.
\end{abstract}


\section{Introduction}

Fast radio bursts (FRBs) with durations $\sim 1$ to 8~ms show dispersive arrival times consistent with a cold plasma and  dispersion measures (DM, the column density of free electrons) too large to be accounted for by the NE2001 model for free electrons  in the Milky Way.   Thus far,  reported FRBs have been detected from
17 distinct sources   \citep[][]{2007Sci...318..777L, 2012MNRAS.425L..71K, 2013Sci...341...53T,  2014ApJ...790..101S,  2014ApJ...792...19B, 2015MNRAS.447..246P, 2015ApJ...799L...5R,  2015arXiv151107746C, 2015Natur.528..523M, 2016Natur.530..453K}.
 They have DM ratios $\DM / \DM_{\rm NE2001, \infty} \sim 1.4$ to 35, where $\DM_{\rm NE2001, \infty}$ is the total integral of the NE2001 Galactic model for the electron density \citep[][]{2002astro.ph..7156C} in the direction of the FRB.

The source of unmodeled  electron-density contributions to FRB DMs has been widely debated in the literature, with explanations ranging from the photospheres of Galactic stars
\citep[][]{2014MNRAS.tmpL...2L}, to host galaxies with negligible contributions from the intergalactic medium \citep[IGM;][]{2016MNRAS.457..232C}, to  the IGM  as the dominant medium
\citep[e.g.][]{2007Sci...318..777L,  2013Sci...341...53T, 2016Natur.530..453K}.  Recent work, however, suggests that  the sources of several FRBs are certain to be extragalactic and that they reside in galaxies.   The repeater FRB121102 \citep[][]{s2016} shows no evidence for an HII region in a  deep VLA image  that could account for the excess DM \citep[][]{scholz2016}.
No HII regions have been seen in any of the other FRB directions, though  FRB010621 at Galactic latitude $b = -4.0^{\circ}$ toward the inner Galaxy ($l = 25.4^{\circ})$ requires further investigation.
The Faraday rotation  measure (RM) of  the high Galactic latitude FRB110523
\citep[][]{2015Natur.528..523M} is consistent with the magnetoionic medium of a spiral host galaxy.
Finally \citet[][]{2016Natur.530..453K} have identified a redshift $z=0.49$ galaxy coincident with afterglow type variability of a source in the field of view of FRB150418, though this association has been questioned by, e.g.,  \citet[][]{wb16,vrm+16}.

Ten out of the 17 known FRB sources produce bursts that show
  asymmetric pulse broadening  with time constant $\taud$ caused by scattering from  small-scale electron-density variations.  The  others show more symmetric pulses that imply  upper bounds on  broadening  comparable to  measured values.   Most of the known FRBs are in directions where the Galactic contribution to $\tau$ is negligible.     To be sure,  the known sample is highly affected by Galactic scattering that prevents detection of fast bursts in directions through the inner Galaxy.

In this paper we define the Galactic $\taud$-\DM\ relation using pulsars and establish that, as a class, FRBs are {\it under scattered} with respect to this relation.   Measurements or upper limits on FRB broadening are therefore notable in two ways:  First, the measured pulse broadening  is much larger than expected from the Milky Way for the directions to FRBs; but, second, the broadening  is  smaller, sometimes significantly so, than would be expected from the $\taud$-\DM\ relation for Galactic pulsars having  the same DM.  Most of the upper limits on FRB scattering are also below the Galactic  $\taud$-\DM\ relation, so this seems to be a general trend. 
We interpret this trend by considering  the physics and geometry of dispersing and scattering electrons that involve the Milky Way, the IGM, a host galaxy, and specific regions within a host galaxy.    The simplest conclusion is that host galaxies dominate pulse broadening and also likely make  significant contributions to measured FRB DMs.

In Sections~\ref{sec:dispersion} and \ref{sec:scattering} we compare distributions of DM and $\taud$ for Galactic pulsars and FRBs.
In Section~\ref{sec:taudm} we analyze the  $\taud$-\DM\ relation for Galactic pulsars
and in Section~\ref{sec:outliers} we discuss pulsar lines of sight that show deficits of scattering
as a prelude to our analysis of FRBs in
Section~\ref{sec:frbs}.
We interpret and summarize our results in  Section~\ref{sec:discussion}.
Appendix~\ref{sec:app} summarizes data from the literature that we have used to determine
the $\tau$-DM distribution of pulsars.
Appendix~\ref{sec:measures} gives relationships between dispersion, scattering, and emission measures used in our analysis.


\section{FRB Dispersion Measures}
\label{sec:dispersion}

The distribution of  DMs for FRBs contains significant information that can be used eventually to constrain  the  distances and environments of FRB sources.   Here we discuss the DM distribution and the next section we discuss FRB scattering. 
Table~\ref{tab:frbs} gives DMs and scattering times for the FRBs we consider along with relevant references.

\begin{deluxetable*}{lrrrrrrrc}
\tablecaption{\label{tab:frbs} Parameters of 17 FRB Sources}
\tablehead{
  \colhead{FRB} & \colhead{$l$} & \colhead{$b$} & \colhead{$\DMfrb$} & \colhead{$\tau$\tablenotemark{a}} & \colhead{DM$_{\rm NE2001}$\tablenotemark{b}} & \colhead{$\tau_{\rm NE2001}$\tablenotemark{c}} & \colhead{$\DMxgal$\tablenotemark{d}} &  \colhead{Ref.} \\
 &  \colhead{(deg)} & \colhead{(deg)} & \colhead{$\rm (pc\ cm^{-3})$} & \colhead{(ms)} & \colhead{$\rm (pc \ cm^{-3})$} & \colhead{($\mu$s)} &  \colhead{$\rm (pc \ cm^{-3})$} & }
\startdata
    010125 &      357 &   $   -20$ &      790 &   $< 19.1$ &      111 &  1.0     &      649 &   1 \\[-2pt]
    010621 &       25 &   $    -4$ &      746 &   $<  5.4$ &      536 &     105.5 &      180 &   2 \\[-2pt]
    010724 &      301 &   $   -42$ &      375 &  $  17.7$ &       45 &    0.1      &      300 &   3 \\  [-2pt]
    090625 &      226 &   $   -60$ &      900 &  $   3.7 $&       32 &    0.05       &      838 &   4 \\[-2pt]
    110220 &       51 &   $   -55$ &      944 &   $ 16$ &       35 &    0.06          &      879 &   5 \\[-2pt]
    110523 &       56 &   $   -38$ &      623 &   $  0.7$ &       44 &    0.1         &      549 &  6  \\[-2pt]
    110626 &      356 &   $   -42$ &      723 &   $<  4$ &       48 &    0.1         &      645 &   5 \\[-2pt]
    110703 &       81 &   $   -59$ &     1104 &   $< 12.3$ &       32 &    0.05    &     1042 &   5 \\[-2pt]
    120127 &       49 &   $   -66$ &      553 &   $<  3.1$ &       32 &    0.05      &      491 &   5 \\[-2pt]
    121002 &      308 &   $   -26$ &     1629 &  $   6.7 $&       74 &    0.3      &     1525 &   4 \\[-2pt]
    121102 &      175 &   $    0$ &      557 &   $<  1.5$ &      188 &   12.4     &      339 &  7  \\[-2pt]
    130626 &        7 &   $    27$ &      952 &    $ 2.9 $&       67 &    0.3 &      856 &   4 \\[-2pt]
    130628 &      226 &   $    31$ &      470 &   $  1.2$ &       53 &    0.2 &      387 &   4 \\[-2pt]
    130729 &      325 &   $    55$ &      861 &   $ 23$ &       31 &    0.06 &      800 &   4 \\[-2pt]
    131104 &      261 &   $   -22$ &      779 &    $12.5$ &       71 &    0.3 &      678 & 8   \\[-2pt]
    140514 &       51 &   $   -55$ &      562 &    $ 5.4 $&       35 &    0.06 &      497 &  9  \\[-2pt]
    150418 &      233 &   $    -3$ &      776 &   $<  3.1$ &      187 &   13.5 &      559 &  10  \\[-2pt]
\enddata
\tablenotetext{a}{Scattering times at 1 GHz}
\tablenotetext{b}{Galactic contribution to the measured DM estimated by integrating the NE2001 model to the edge of the Galaxy.}
\tablenotetext{c}{Galactic contribution to the pulse broadening estimated by the NE2001 model.}
\tablenotetext{d}{Extragalactic contribution to the measured DM: $\DMxgal = \DMfrb$ - DM$_{\rm NE2001}$ - DM$_{\rm halo}$, where
DM$_{\rm halo}$ = 30 pc cm$^{-3}$.}
\tablenotetext{}
{References:
1 - \cite{2014ApJ...792...19B},
2 - \cite{2012MNRAS.425L..71K},
3 - \cite{2007Sci...318..777L},
4 - \cite{2015arXiv151107746C},
5 - \cite{2013Sci...341...53T},
6 - \cite{2015Natur.528..523M},
7 - \cite{2014ApJ...790..101S},
8 - \cite{2015ApJ...799L...5R},
9 - \cite{2015MNRAS.447..246P},
10 - \cite{2016Natur.530..453K} }
\end{deluxetable*}

\begin{figure}[t!]
\begin{center}
\includegraphics[scale=0.45]{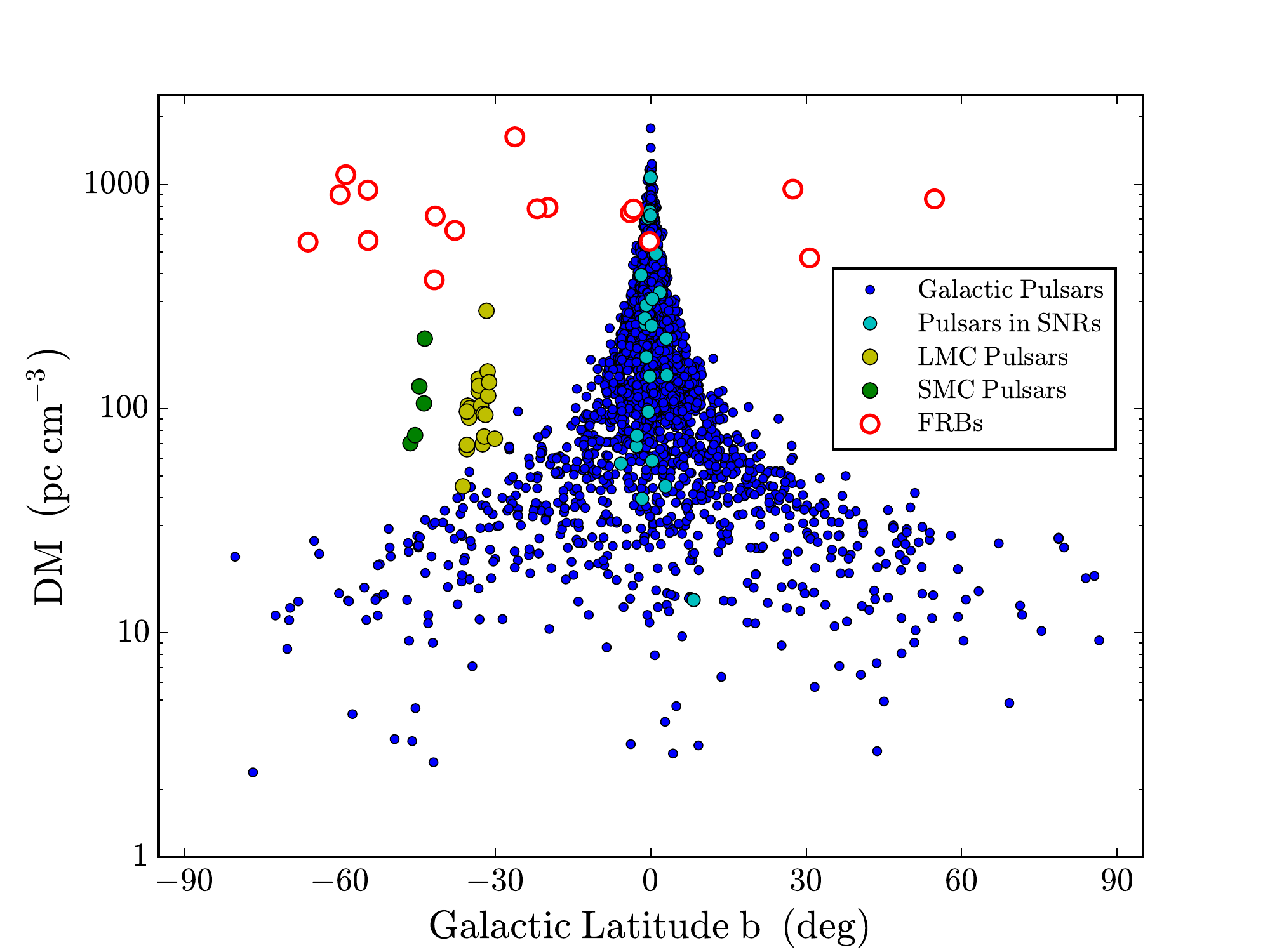}
\caption{Dispersion measures plotted against Galactic latitude.  Different symbols are used  for Galactic pulsars (2422 objects), Galactic pulsars associated with supernova remnants (27), pulsars in the Large Magellanic Cloud (LMC, 21) and Small Magellanic Cloud  (SMC, 5), and FRBs (17). DM measurements and
pulsar associations were obtained from
\citet[][http://www.atnf.csiro.au/research/pulsar/psrcat]{2005AJ....129.1993M} .
%
The two FRBs  with similar DMs near $b=-20^{\circ}$ (FRB010125 and FRB131104) are not close on the sky because their longitudes are very different;  nor are the two FRBs near  $b=-3^{\circ}$ (FRB010621 and FRB150418).
The FRB nearest to $b=0^{\circ}$ is the repeater  FRB121102. Its DM is actually significantly larger than the Galactic contribution because it is in the Galactic anticenter direction, whereas the largest pulsar DMs are near $b = 0^{\circ}$   toward the Galactic center.
\label{fig:dm_vs_b_psrs_frbs}
}
\end{center}
\end{figure}

In Figure~\ref{fig:dm_vs_b_psrs_frbs} we show FRB DMs plotted against Galactic latitude $b$ 
and for comparison we show DMs for  pulsars in the Milky Way and in the Magellanic Clouds (LMC, SMC). 
Several conclusions can be made from the figure.

First,  the DMs of all FRBs with $\vert b \vert > 10^{\circ}$  are much larger than the outer envelope of the distribution of DMs for Galactic pulsars, which approximately follows a $\csc \vert b \vert$ dependence.
An extragalactic population of FRBs would appear just this way if the total DM includes  a large extragalactic component.

Second, the total range of  FRB DMs  is no greater than that seen  from Galactic
pulsars. FRB DMs are unextraordinary column densities that could be provided by
 dwarf galaxies in some cases.  For example, the smallest FRB DM is only 37\% larger than the largest  DM seen for a pulsar (J0537$-$69) in the LMC (after correction for Galactic contributions to the DMs using the NE2001 model and also a halo correction for the FRB DM,
 as discussed below in Section~\ref{sec:frbs}).
The largest FRB DMs correspond to integration through a large portion of a galaxy disk or through a galactic center like that of  the Milky Way.   Ionized gas in galaxies therefore is a plausible source for some or most of the extragalactic part of DM.

Third, for $\vert b \vert > 10^{\circ}$, there is a significant gap
$\Delta\DM \sim 300$~pc~cm$^{-3}$  between the
the smallest FRB DM  (375~pc~cm$^{-3}$ for FRB010724) and the largest pulsar DM at the same latitude.
One possibility is that the IGM is responsible for the  FRB's non-Galactic DM and the gap is 
 the minimum DM that corresponds to the path length through   a volume large enough to contain an emitting source in the appropriate time frame. 
If so, a homogenous and isotropic population of objects would show a DM  probability density function (PDF)  that scales
(in the mean) as $\DM^2$ out to a volume-limited maximum for a standard-candle population;  the corresponding cumulative distribution function (CDF) would scale as $\DM^3$. 
If DMs are dominated by the IGM,  continued discoveries should yield  smaller DMs than in the current FRB sample of 17 objects (as of 2016 March 1).  FRBs out to 100 Mpc, for example,  would show IGM contributions of only 50~pc~cm$^{-3}$.

\begin{figure}[t!]
\begin{center}
\includegraphics[scale=0.45]{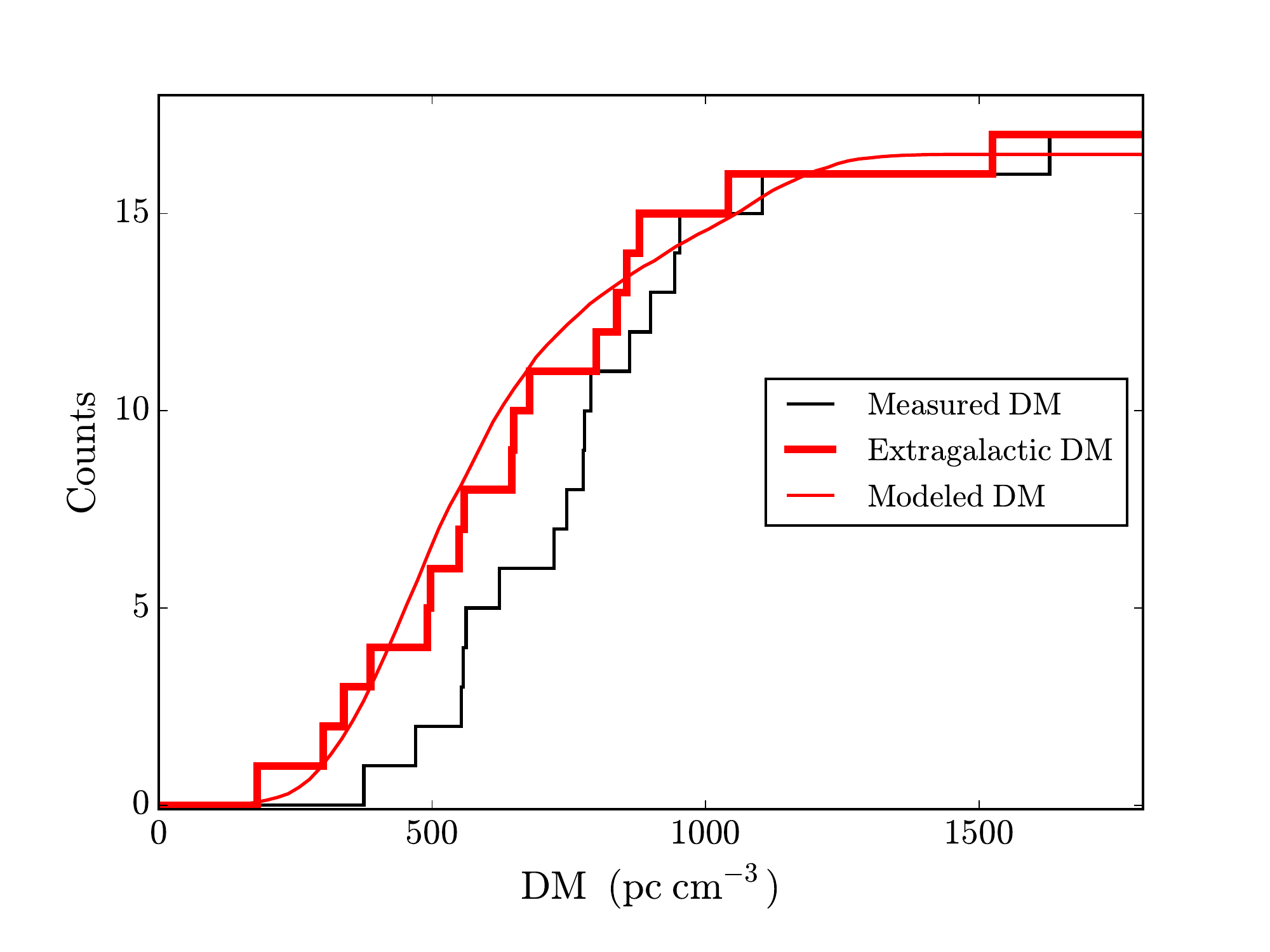}
\caption{Cumulative density functions  (CDFs) of DM for the 17 known FRBs:
measured DMs  (black);  extragalactic portions of DMs (heavy red line);
model DM (light red) for  a compact spherical source region in an ellipsoid with a 3:1 axial ratio.
The extragalactic DM contributions are found by subtracting  the Galactic portion equal to  the  NE2001 estimate for the direction of the FRB added to a halo contribution of
30~pc~cm$^{-3}$ \citep[][]{2015MNRAS.451.4277D}.
\label{fig:cdf_data_model}
}
\end{center}
\end{figure}

To explore this further, we show in 
Figure~\ref{fig:cdf_data_model} the CDFs for FRB DMs as measured and after subtraction of the Galactic contribution,
estimated from the disk contribution from the  NE2001 model for the direction of the FRB added to a halo contribution of
30~pc~cm$^{-3}$ \citep[][]{2015MNRAS.451.4277D}.    The role of selection effects in the DM distributions is not known at present, so in addition to small-number statistics, the shape may be biased.  At face value, however, the CDFs (either before or after correction for the Galactic contribution) are not consistent with DMs that are dominated by the IGM. 

An alternative interpretation is that the DM gap mentioned above, 
  combined with the range of DMs extending up  to $\sim 1600$~pc~cm$^{-3}$,  may imply that FRB sources reside in host-galaxy regions  with large electron densities, such as the interiors of  supernova remnants, HII regions, and the centers of galaxies.   Otherwise FRB sources at arbitrary sites within spiral galaxies would show small as well as large DMs related to the orientation of the galaxy. If FRB sources are intermixed with the free electrons responsible for the dispersion, small DMs would be more probable.

\begin{figure}[t!]
\begin{center}
\includegraphics[scale=0.35]{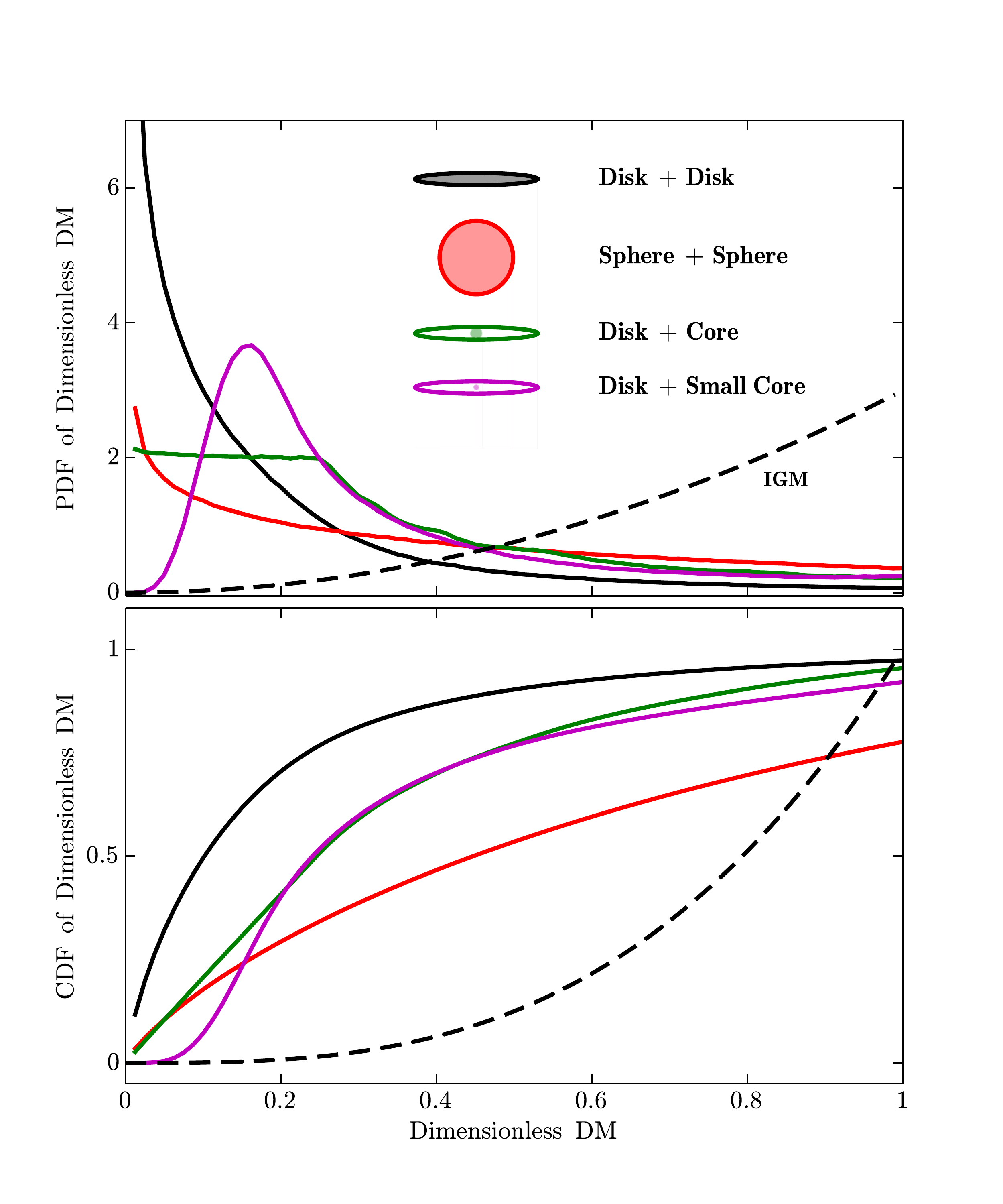}
\caption{(Top) Probability density functions (PDFs) of DM for schematic distributions of free electrons and sources of discrete pulses.
The PDFs  have been calculated from 500 different  observers in random directions and at  a distance much larger than the source population.    The legend gives the shapes of the free electron distribution and the FRB source distribution.     Electrons have a
filled  distribution  with its characteristic size depicted as a solid line contour.  The source distribution is shown as a color-filled shape.
Three out of the four distributions are disks seen edge on with a 10:1 axial ratio.  The fourth is spherical.
The dashed line shows the PDF expected for the IGM, which is a spherical distribution with the observer at the center.
Note that none of the PDFs shown take into account selection effects in FRB surveys.
\label{fig:schematic_dm_cases}
(Bottom) 
Cumulative density functions  (CDFs) of DM for the schematic distributions shown in Figure~\ref{fig:schematic_dm_cases}.
\label{fig:schematic_dm_cases_cdfs}
}
\end{center}
\end{figure}

To illustrate this interpretation we show PDFs of the dimensionless DM for some simple geometries in Figure~\ref{fig:schematic_dm_cases}
(top panel).  Each case has a different  shape for the spatial distribution of free electrons and  for FRB sources.
 Curves were   generated by Monte Carlo using 2000 objects
for each of 500 isotropic observer directions, with the observer at a large distance from the population.   To dimensionalize the DM values, they must be multipled by a length scale and by a peak electron density.  The length scale could be the radius of a disk galaxy or it could be the size of a supernova remnant  (SNR) or galactic center.
The heavy lines delineate the shape and size of the electron distribution (e.g. the 1/e scale) while the filled region represents the distribution of FRB sources.   The top case (`Disk + Disk') is for a disk galaxy, which is shown in a side view. The top two cases in the figure have identical electron and source distributions.   The bottom two have source distributions that are smaller than the electron distributions.

\begin{figure}[t!]
\begin{center}
\includegraphics[scale=0.45]{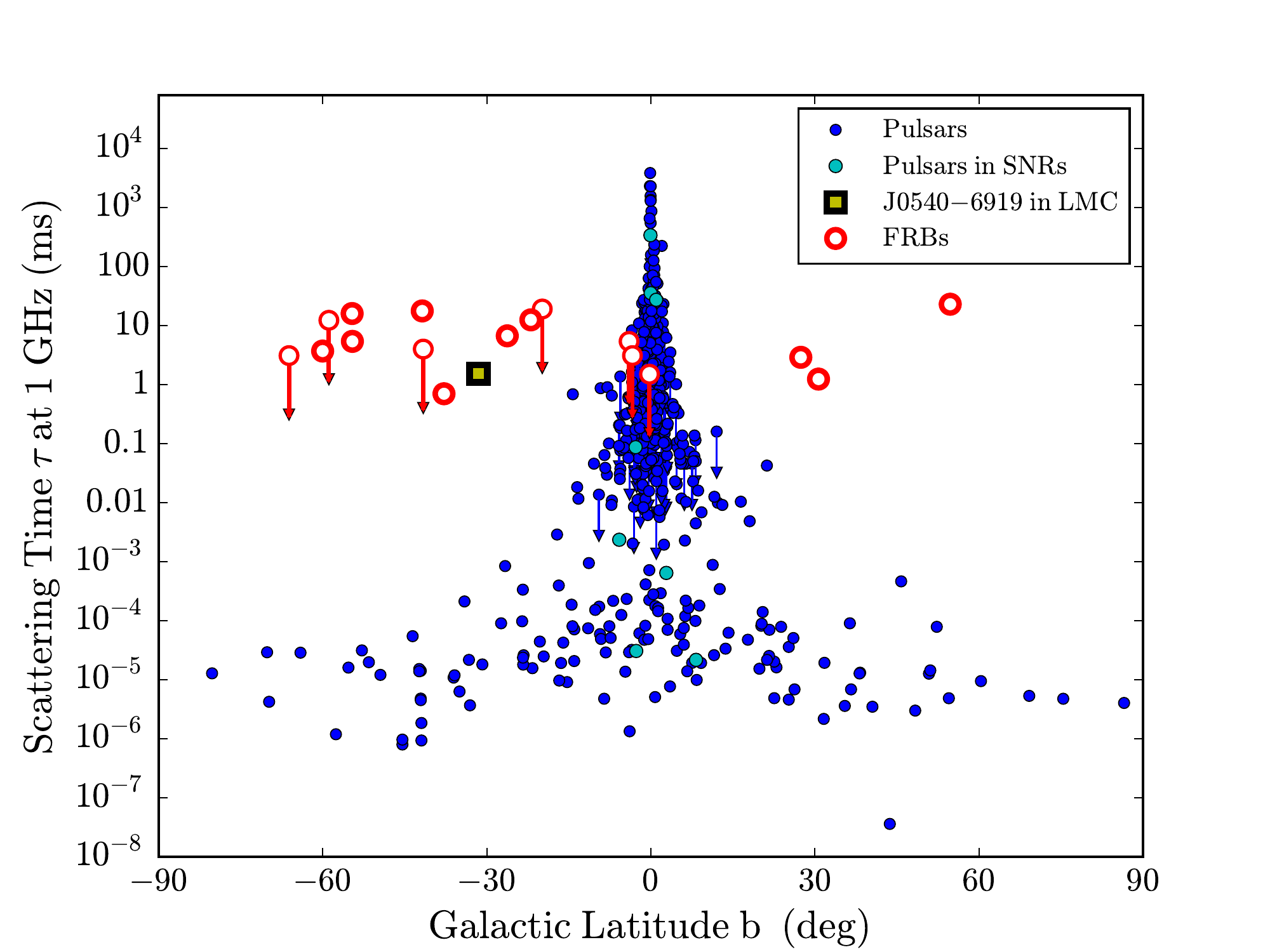}
\caption{Pulse broadening times for pulsars and FRBs at 1~GHz plotted against Galactic latitude.
There are 421 pulsar measurements and 93 upper limits on $\taud$ compared to 10 FRB measurements and 7 upper limits.
\label{fig:tau_vs_b_psrs_frbs}
}
\end{center}
\end{figure}

   When the two distributions are equal in size in at least one dimension, the resulting PDF maximizes at $\DM = 0$ whereas a source distribution that is smaller than the electron distribution in all dimensions yields a peak at non-zero DM.  This is demonstrated for the case
labeled `Disk + Small Core' but also results when a spherical source distribution is smaller than a spherical electron distribution.    The peak of the DM distribution shifts to larger DM in proportion to the ratio of radii of the two spherical two distributions.   FRB sources in different galaxies would show a combined DM distribution that is a superposition of single-galaxy PDFs and there would be no correlation of DM with distance.

Figure~\ref{fig:schematic_dm_cases_cdfs} (bottom) shows cumulative distributions (CDFs) whose shapes display 
  differences in convexity.  
We overplot (light red line) a representative fit to the CDF for extragalactic DMs using an ellipsoidal electron-density distribution with axes in proportion to 1:1:0.35 and a spherical source distribution of relative size 0.1.    The detailed parameters are not unique but the FRB distribution seems to require a model that is generically similar to this example.   The  concave distribution expected from an IGM-dominated model ($\propto \DM^3$) cannot fit the data unless it is heavily modified by selection effects.   A detailed study is deferred to another paper.
%

\section{FRB Scattering}
\label{sec:scattering}

Figure~\ref{fig:tau_vs_b_psrs_frbs} shows scattering times $\tau$ at 1~GHz vs. Galactic latitude.  The data used in the figure are
based on pulse broadening and scintillating bandwidth measurements scaled to 1~GHz; they are summarized in 
Appendix~\ref{sec:app}.

Pulsar scattering times span more than ten orders of magnitude. The scattering times of FRBs, like their DMs,  are  also within the range spanned by pulsars but are many orders of magnitude larger than pulsars at similar Galactic latitudes, in most cases.  The large values of $\tau$ signify either that FRBs are Galactic with large column densities of unmodeled free electrons along their lines of sight or that they are extragalactic.    Recent evidence supports an extragalactic origin for two FRBs based on scintillations and Faraday rotation of FRB110523  \citep[][]{2015Natur.528..523M} and from the absence of a radio-emitting HII region along the line of sight to FRB121102 \citep[e.g.][]{scholz2016}.

Of course some of the FRBs could still be Galactic.
In the following we assume that the 17 FRBs included in Table~\ref{tab:frbs} are extragalactic and assess the nature of FRB scattering by comparing to  Galactic scattering of pulsars.

\section{The  $\taud$-\DM\ Relation for Galactic Pulsars}
\label{sec:taudm}

\begin{figure}[t!]
\begin{center}
\includegraphics[scale=0.45]{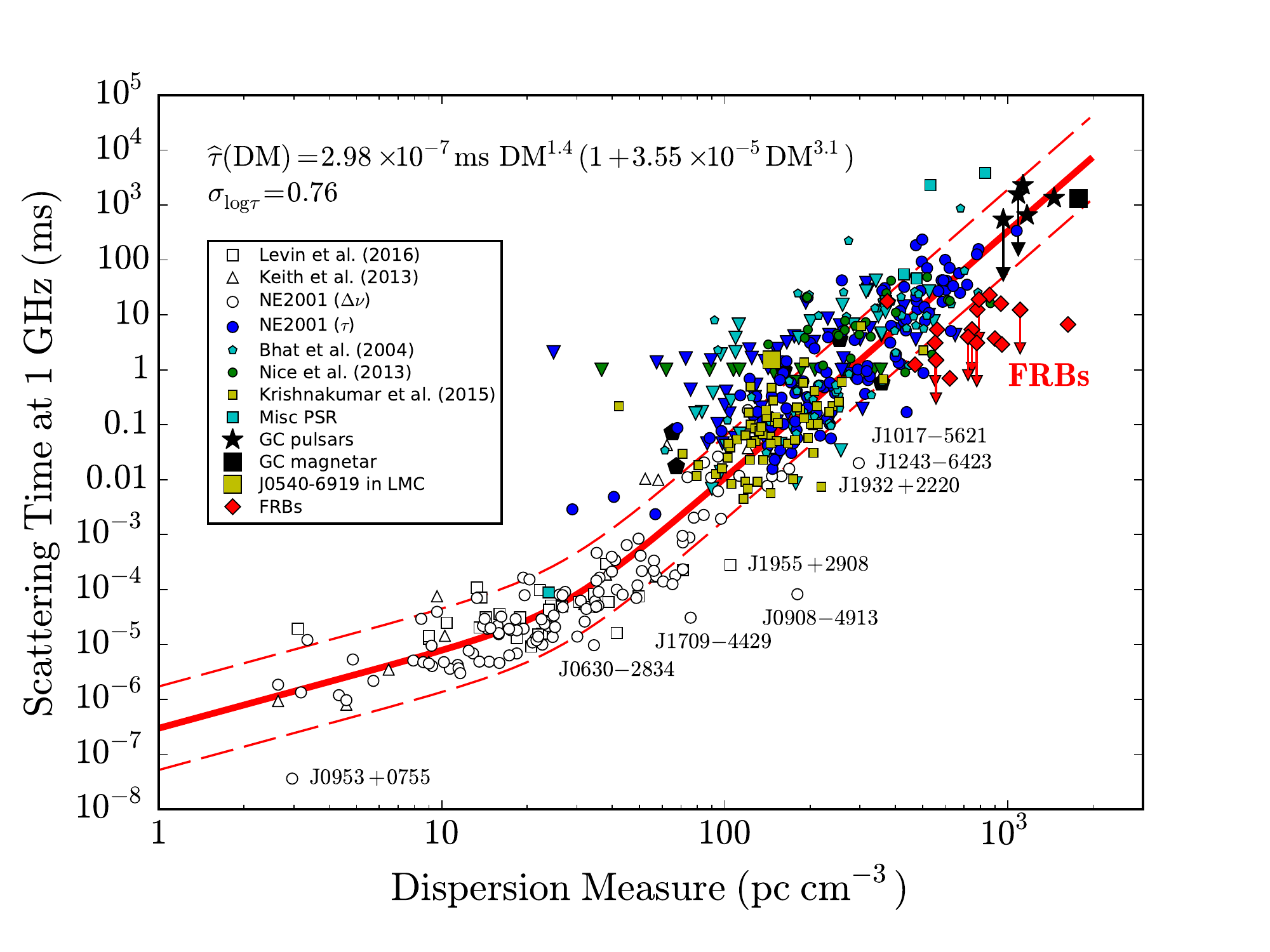}
\caption{The $\tau$-$\DM$ plane for Galactic pulsars and fast radio bursts.  The legend gives the color codes for different kinds of objects or different data sources.   Upper limits are denoted by downward going triangles or with arrows for FRBs and pulsars near the Galactic center.
\label{fig:tau-dm}
}
\end{center}
\end{figure}

Figure~\ref{fig:tau-dm}  shows  $\taud$ plotted against \DM\ for 531 lines of sight, including
421 measurements and 93 upper limits on pulsars and a magnetar and 17 FRB values  (10 measurements and 7 upper limits).    Data
types and sources are summarized in Appendix~\ref{sec:app}.

\begin{figure}[t!]
\begin{center}
\includegraphics[scale=0.45]{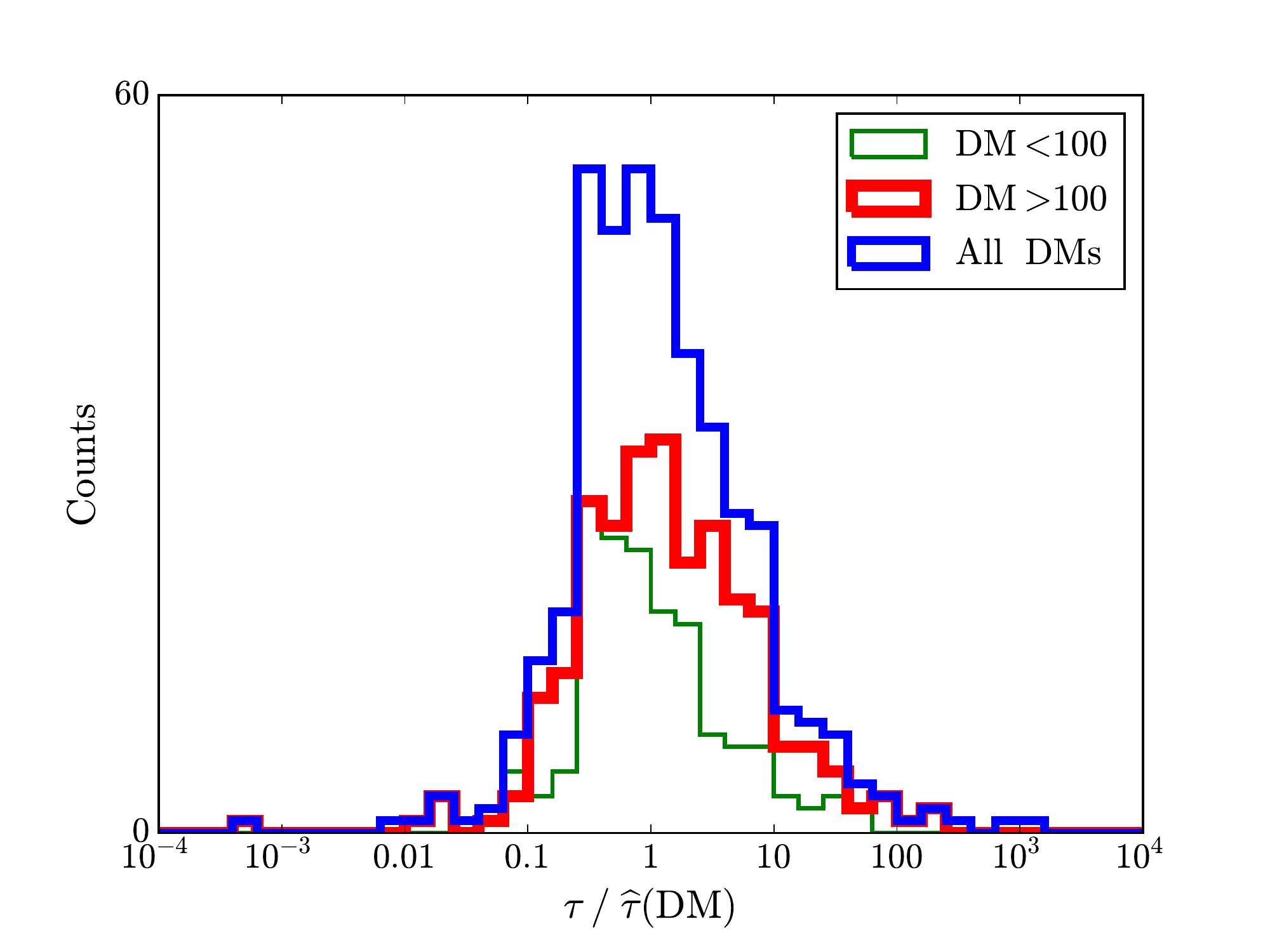}
\caption{
Histograms of the residuals between measured $\log\tau$ and the model $\log\taudhat$ for
pulsars only.  Histograms are shown for pulsars with $\DM < 100$~pc~cm$^{-3}$ and
$\DM > 100$~pc~cm$^{-3}$ as well as for all DMs.
\label{fig:dloghist}
}
\end{center}
\end{figure}

Despite a significant vertical  spread in $\taud$ at any given \DM\, there is a clear trend of increasing $\tau$ with \DM. To quantify the trend for Galactic sources (excluding FRBs and a pulsar in the LMC)  we fit a Gaussian model
to  $\log \taud$ with a standard deviation $\sigma_{\log\taud}$  and  a mean scattering time of  the form  \citep[][]{1997MNRAS.290..260R}
\be
\taudhat(\DM) = A\times\DM^a (1 + B\times\DM^b).
\label{eq:tau-dm-model}
\ee
The RMS variation $\sigma_{\log\taud}$  is assumed to be  independent of \DM\ and is also a search parameter.   The likelihood function for measurements was taken to be the product of factors involving a Gaussian probability density
function (PDF),
${\cal L}_j = (2\pi\sigma_{\log\taud}^2 )^{-1/2} \exp[-(\log({\taud}_j / \taudhat)^2 / 2 \sigma_{\log\taud}^2]$, for the $j^{\rm th}$
pulsar.  For upper limits we used the CDF
${\cal L}_j = (1/2)[1 + {\rm erf}( \log({\taud}_j / \taudhat) / \sqrt{2}\,\sigma_{\log\taud}]$.

The maximum likelihood solution, obtained from a grid search, is
\be
A &=& 2.98\times10^{-7}~{\rm ms},
\quad
B = 3.55\times 10^{-5},
\nonumber \\
a &=& 1.4,
\quad
b = 3.1,
\quad
\sigma_{\log\taud} = 0.76,
\label{eq:tau-dm-model-parameters}
\ee
with roughly 5\% errors on each parameter.
Figure~\ref{fig:tau-dm} shows $\taudhat(\DM)$ as a solid red line and
${\rm dex} \left[{\log \taudhat(\DM) \pm \sigma_{\log\taud}}\right]$ as dashed red lines.

The distribution of pulsar scattering times relative to the best fit trend appears to have positive skewness. Figure~\ref{fig:dloghist} shows  histograms of $\log \tau / \log\taudhat$
for $\DM < 100$~pc~cm$^{-3}$, for  $\DM > 100$~pc~cm$^{-3}$, and for all DMs.   The low-DM histogram is clearly skewed while the high-DM histogram is also skewed but less so.   The skewness is consistent with scattering along  the  lines of sight  typically being dominated by a small number of scattering regions.  While longer lines of sight encounter more regions, yielding a more symmetric distribution,  the chances are also higher for encountering a region with stronger scattering that dominates $\tau$ \citep[e.g.][]{1991Natur.354..121C}.

\section{Pulsar Outliers}
\label{sec:outliers}

Figures~\ref{fig:tau-dm} and \ref{fig:dloghist} show
clear outliers  from the $\taud$-\DM\ relation, both above and below the fit, in some cases by two or more times the standard deviation $\sigma_{\log\taud}$.     Excess scattering is easily produced by HII regions along some lines of sight, with increasing likelihood at large DMs that usually correspond to larger distances.   Some of the HII `clumps' in the NE2001 model were in fact identified  this way.   As specific examples, the two largest scattering times are the cyan squares in Figure~\ref{fig:tau-dm} at $\DM \sim 532$ and 830~pc~cm$^{-3}$ for pulsars J1841$-$0500 and J1550$-$5418.  The first of these is near the SNR Kes~73 but not close enough to attribute the scattering to the SNR. However, the pulsar location is near ISM wisps that may provide enhanced scattering somewhere along the $\sim 7$~kpc path length \citep[][]{2012ApJ...746...63C}.   The second object is a magnetar coincident on the sky
and possibly associated with the SNR G327.24$-$0.13  \citep[][]{2007ApJ...666L..93C}.   The SNR appears to be responsible for the enhanced scattering.

Another pulsar with excess scattering is the young pulsar J0540$-$6919 (B0540$-$69) in the LMC with $\DM \sim 146$~pc~cm$^{-3}$
and $\tau \sim 1.5$~ms at 1~GHz, about 5000 times larger than the NE2001 prediction.   The scattering almost certainly comes from the LMC\footnote{The Galactic halo extends far beyond the LMC and in total has an emission measure $\EM\sim 5 \times 10^{-3}$~pc~cm$^{-6}$
 \citep[][]{2012ApJ...756L...8G}, which puts an upper bound on the scattering measure and thus on the  pulse broadening, $\tau_{\rm halo} \lesssim 50$~ns. See Appendix~\ref{sec:measures} and Eq. 9 of \citet[][]{2002astro.ph..7156C}.} and evidently   is strong enough to overcome the geometric deleveraging of $\tau$ by the proximity of the scattering electrons to the source.

There are also significant {\it deficits} in $\taud$  for some pulsars, a few of which are labeled in the figure.
We discuss these in terms of a toy model for scattering that can account for the deficits and will guide our analysis of FRB lines of sight that also show deficits.


\subsection{Simple Scattering Model for Pulsar Lines of Sight}
\label{sec:psrsimple}

We calculate $\tau$ by integrating the mean-square scattering angle per unit distance $\eta(s)$ over a path length $d$,
\be
2c\taud = \int_0^d ds\, \eta(s) s(1-s/d),
\label{eq:2ctau}
\ee
where the line of sight weighting enforces  that
propagation paths from the source at $s=0$ reach the observer at $s=d$
\citep[][Appendix A]{1985MNRAS.213..591B}.
Here and throughout the paper we assume  angles are very small compared to a radian.
Eq.~\ref{eq:2ctau} also assumes  scattering is homogenous across an infinite transverse plane at each $s$.   If the scattering region is truncated or otherwise inhomogeneous, the pulse broadening is altered \citep[][]{2001ApJ...549..997C}.  Screen truncation may be relevant to FRB scattering and is discussed later.

The mean-square scattering angle can be related to other bulk quantities that characterize electron density fluctuations (Appendix~\ref{sec:measures}).  We use a Kolmogorov wavenumber spectrum \citep[e.g.][]{1995ApJ...443..209A}
with wavenumber cutoffs corresponding to an inner scale ($\linner$) and an outer scale ($\louter$).
Then $\eta = h_\lambda \Ftilde n_e^2$ where  $n_e$ is the volume-average electron density
and  $h_{\lambda} =   \Gamma(7/6) \lambda^4 r_e^2$;  $\lambda$ is the wavelength, $\Gamma$ is the gamma function, and $r_e$ is the classical electron radius \citep[c.f. Appendix A of][]{1998ApJ...507..846C}.
The  `fluctuation' parameter $\Ftilde = \zeta \epsilon^2 / \ff (\linner \louter^2)^{1/3}$ involves the
inner scale because the smallest scales scatter the radiation the most while the outer scale is part of the relation between the variance of the total density  and the squared-mean density,
$n_e^2$.
Other quantities parameterize  the  fractional density variance $\epsilon^2$ inside small HII clouds; cloud-to-cloud variations are described by the dimensionless second moment $\zeta$ and the volume filling factor $\ff$.

For a homogeneous medium with constant $\eta$, the broadening time $2c\tau = \eta d^2 / 6$ can be related to
$\DM =  n_e d$ as $2c\tau =  h_\lambda \Ftilde \DM^2/6$, which gives
\be
\tau = 0.745~{\rm s}\, \zeta \epsilon^2 \nu^{-4} \DM_{1000}^2
	\left[ \frac{l_i}{10^3~\rm km} \left(\frac{l_o}{100~\rm pc}\right)^2\right]^{-1/3}
\ee
for $\nu$ in GHz, DM in units of $10^3$~pc~cm$^{-3}$, and fiducial inner and outer scales of $10^3$~km and 100~pc,
respectively, that are like those in the NE2001 model.
These values are also of order the same that were found from a detailed study of pulsar broadening for  the pulsar J1644$-$4559
\citep[][]{2009MNRAS.395.1391R}.
 The nominal broadening time matches those in
Figure~\ref{fig:tau-dm} for $\DM = 10^3$~pc~cm$^{-3}$.

The fit $\taudhat(\DM)$  in Figure~\ref{fig:tau-dm} has segments
$\propto \DM^{1.4}$ and $\propto \DM^{4.4}$ that differ markedly from this $\tau \propto \DM^2$ scaling.  This is evidently due to  strong  inhomogeneities of the ISM, leading to large differences  in $\Ftilde$ across the Galaxy.
The NE2001 model includes components between which $\linner^{1/3} \Ftilde$ varies by a factor $\gtrsim 500$, with the largest values in the inner Galaxy and in spiral arms \citep[c.f. Table 3 in][]{2002astro.ph..7156C}.

\begin{figure}[t!]
\centering
\includegraphics[trim = 30mm 35mm 15mm 35mm, clip, scale=0.575]{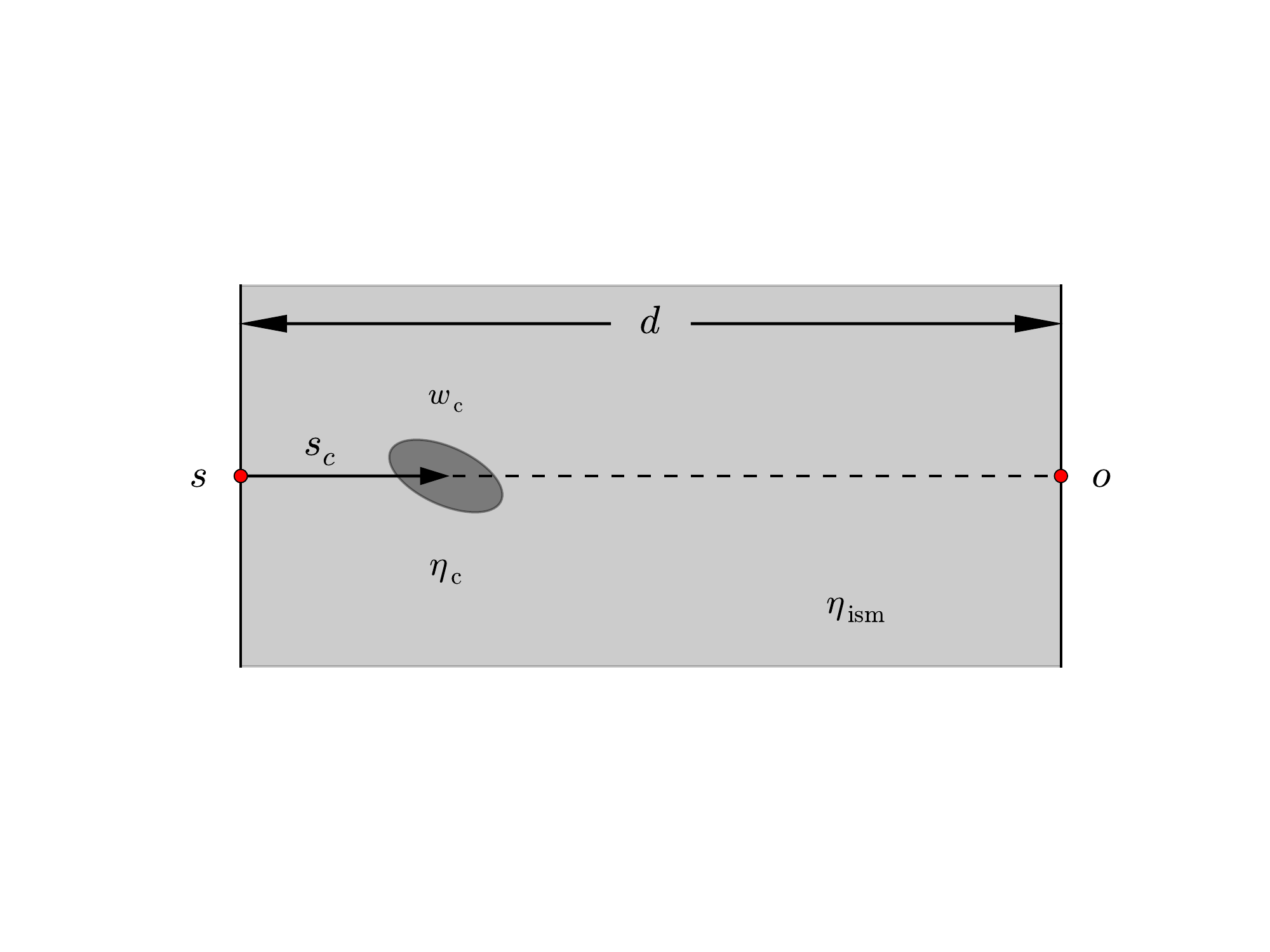}
\caption{
Simple model for Galactic scattering from a background ISM with constant angular scattering variance per unit distance, $\eta_{\rm ism}$, combined with a clump of depth $w_c$  at distance $s_c$ from a source and having a  different angular scattering  variance per unit length, $\eta_c$.
The source-observer distance is $d$.
\label{fig:geom_galactic}
}
\vspace {0.5in}
\end{figure}

For pulsar scattering,
we adopt a simple model (Figure~\ref{fig:geom_galactic}) that includes  a constant background level of scattering $\etaism$ combined with scattering $\eta_c$ from a clump  at a distance  $s_c$ from the pulsar.   The clump has a line-of-sight extent   $w_c \ll d$.
We write  $\eta(s) = \etaism  + \eta_c w_c \delta(s-s_c)$, where $\delta$ is the Dirac delta functional, to obtain
\be
2c\taud &=&  \etaism d^2 / 6 + \etac \wc \sc  (1 - \sc/ d)
\label{eq:tauism}
\\
&=& h_\lambda[  \Ftism {\DMism}^2 / 6 + \Ftc (\sc/\wc) (1-\sc/d) ].
\label{eq:tauism2}
\ee

We compare  two cases, one with  (wc) and one without a clump   along the line of sight,
that  yield the same total \DM\ but correspond to different pulsar distances.
The clump's contribution to \DM\ is $\dDMc$ and the distances are related by
$\dwc = d - \delta\DM_c / \nezero \le d$  for a constant
volume-averaged electron density $\nezero$ for the background ISM.
The ratio of scattering times is
\be
\frac{\tau_{\rm wc}}{\tau} &=& \left(1 - \frac{\delta\DM_c}{\DM} \right)^2
\nonumber \\
	&&+6
	\left(\frac{\delta\DM_c}{\DM} \right)^2
	\left(\frac{ \Ftc}{ \Ftism} \right)
	\left(\frac{\sc}{\wc} \right)
	\left(1 - \frac{\sc}{\dwc} \right)	.
\label{eq:tauratio2}
\ee
This equation can be expressed in terms of  the dimensionless quantities $\rc \equiv \dDMc / DM$
and $\xic \equiv (\Ftc / \Ftism)(\sc /\wc)(1-\sc/\dwc)$,
\be
\tau_{\rm wc} / \tau
= \left(1 - \rc \right)^2 +
	6 \rc^2 \xic
\label{eq:tauratio3}
\ee
where  we define the clump and interstellar portions of the \DM\ as
$\dDMc = \rc \DM$ and $\DMism = (1-\rc)\DM$.

The general solution to the quadratic equation \ref{eq:tauratio3} for $\rc < 1$ is
\be
\rc = \frac{\displaystyle 1 - \sqrt{\tau_{\rm wc} / \tau -6\xic (1-\tau_{\rm wc} / \tau)}}{1 + 6\xic}.
\label{eq:rcsolution}
\ee
  To influence the solution on $\rc$,
$\xic \gtrsim (\tau_{\rm wc} / \tau) / 3 ( 1 - \tau_{\rm wc} / \tau)$ is needed. 
Clumps at average positions along the line of sight ($\sc \sim \dwc / 2$) can satisfy these conditions and lead to the strong excess pulse broadening seen on many of the large-DM cases in
Figure~\ref{fig:tau-dm}. 

\begin{figure}[t!]
\centering
\includegraphics[scale=0.45]{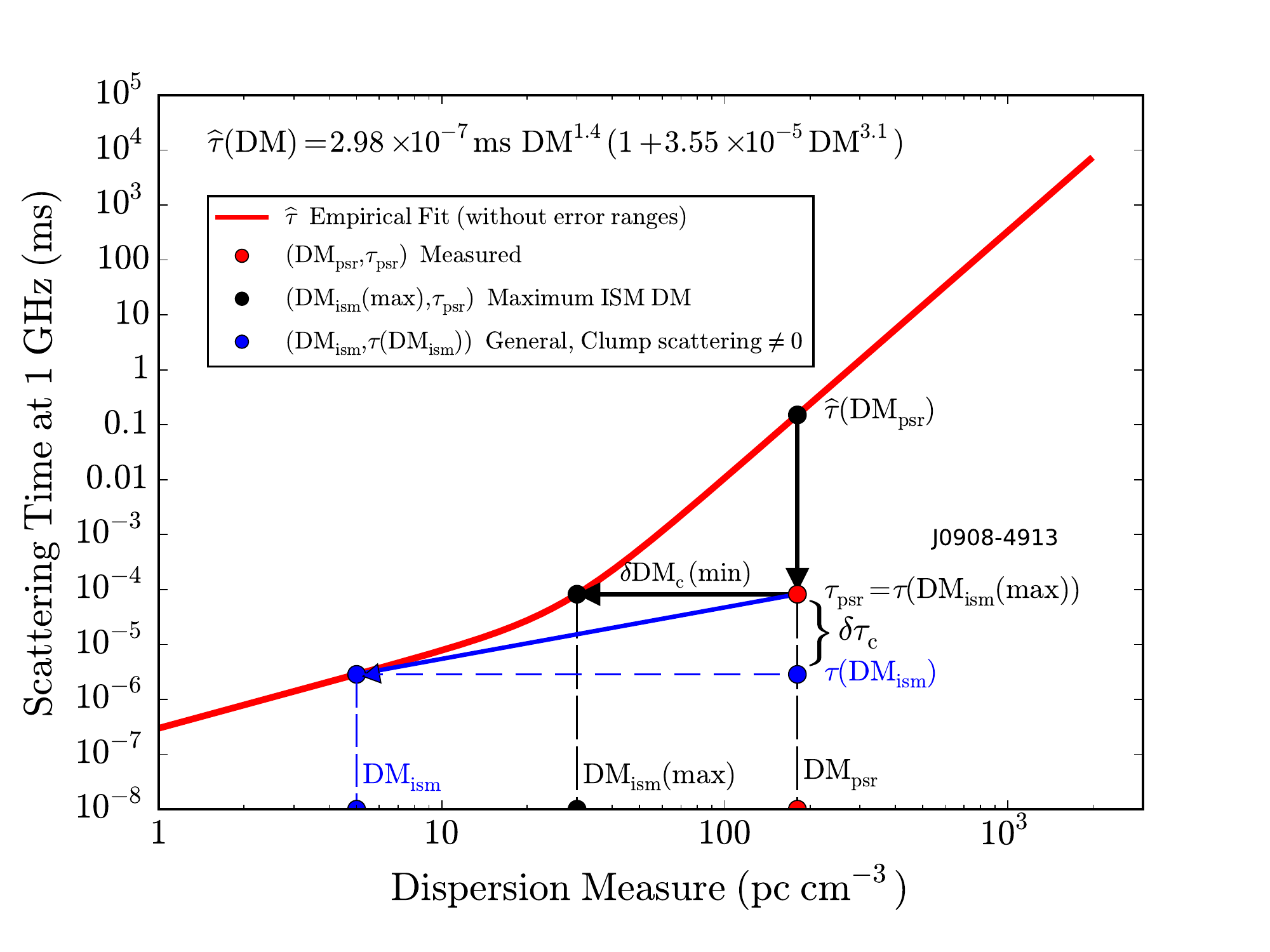}
\caption{
Schematic presentation of the $\tau$-$\DM$ plane and how a pulsar line of sight with a deficit in pulse broadening can be interpreted
(in this case for J0908$-$4913).  The red line shows the empirical fit of Eq.~\ref{eq:tau-dm-model} and various points are labeled in the legend. The minimum required contribution to  DM from a clump
along the line of sight is $\dDMc(\rm min)$ and corresponds to the maximum for the distributed ISM of
$\DMism(\rm max)$.
\label{fig:taudm_schematic}
}
\end{figure}

Excess ionized gas near a pulsar, such as in a bow shock nebula or SNR,  is more likely to perturb only DM unless 
$\Ftc$ is very large. Inspection of Eq.~\ref{eq:tauratio2} or \ref{eq:tauratio3} indicates that   a clump that alters only the DM and does not scatter ($\Ftc \propto \xic= 0$) produces a deficit in the pulse broadening time, $\tau_{\rm wc} / \tau < 1$.
The solution  $\rc= 1 - \sqrt{\tau_{\rm wc} / \tau }$ applies in this case. 
This solution is illustrated schematically in Figure~\ref{fig:taudm_schematic} along with another where the clump contributes to the
scattering but not enough to increase $\tau_{\rm wc}/\tau > 1$.


 \subsection{Pulsars with Scattering Deficits}

 Pulsars that show significantly less scattering than the
 mean $\tau$-DM trend are instructive for our analysis of FRBs and their scattering deficits.  A more detailed analysis of these and other pulsar lines of sight is deferred to another paper as part of the overall development of a new model for free electrons in the Galaxy that will replace the NE2001 model.
%

We offer  three explanations for the pulsars labeled in Figure~\ref{fig:tau-dm} that have scattering deficits:
(1) ionized gas that has intrinsically small fluctuations, as quantified by  the fluctuation parameter $\Ftilde$;
(2) higher-density gas affected by the pulsar itself, such as a pulsar wind nebula (PWN), that enhances the DM but not necessarily
$\tau$ because $\xic \ll 1$; note that $\xic$ can be small due to small $\Ftilde$ or from geometrical effects;
and
(3) higher-density gas in a  region relatively close to but not directly affected by the pulsar;  a spiral arm or supernova shock in the foreground of a pulsar would increase its DM but also would be geometrically disfavored to increase $\tau$.  Examples of all three of these cases can be found. 

An example of the first is
J0953+0755 (B0950+08),  a nearby pulsar \citep[$d =262\pm 5$~pc,][]{2002ApJ...571..906B} viewed through the local super bubble and local hot bubble that evidently have small values of $\Ftilde$ because these regions have  smaller density fluctuations
\citep[][]{1992Natur.360..137P}.

Two objects associated with PWNs are
 J0908$-$4913 (B0906$-$49),  the most under scattered pulsar in Figure~\ref{fig:tau-dm}, and J1709$-$4429 (B1706$-$44).
 If these objects were replotted in Figure~\ref{fig:tau-dm} using only the interstellar portions of their  DMs,  they could shift to the left
 into the normal range of  $\tau$-DM values (c.f. Figure~\ref{fig:taudm_schematic}).
  The minimum required shift for J0908$-$4913 calculated from Eq.~\ref{eq:rcsolution} with $\xic = 0$ is $\sim 130$~pc~cm$^{-3}$  ($\sim 73$\% of its measured DM) and $\sim 45$~pc~cm$^{-3}$ for  J1709$-$4428 (60\%).    These are substantial contributions  that require  electron densities
$\sim 45 ({\rm 1~pc} / \wc) $ and $\sim 130 ({\rm 1~pc} / \wc) $ ~cm$^{-3}$ for the two objects, respectively,  for PWN depths $\wc$.

 J0908$-$4913  has a PWN with a minimum density  $n \gtrsim 2~\rm cm^{-3}$  \citep{gsf98} and is
 also seen through the Gum nebula \citep{pgs15}.  Together these regions could contribute 100~pc~cm$^{-3}$ or more to the total DM.
H$\alpha$ emission as seen in the Southern H$\alpha$ Sky Survey Atlas \citep[SHASSA,][]{gmr01} toward J1709$-$4429
 has a minimum emission measure  $ \EM = 80$~pc~cm$^{-6}$ that is much larger than the NE2001 value of $\rm \widehat{EM} = 0.1-2.2$~pc~cm$^{-6}$.   For a depth $L = 1$~to~10~pc ,  $\DM_{\rm H\alpha} \sim  \sqrt{L\times \EM} \sim 9$ to 28~pc~cm$^{-3}$.
 While there is considerable latitude in these numbers, they are not inconsistent with those required to bring the lines of sight into consistency with the general $\tau$-DM relation.

 The remaining pulsars with scattering deficits appear to require the third explanation.
J0630$-$2834 (B0628$-$28) has a parallax distance  of  $0.33^{+0.05}_{-0.01}$~kpc \citep{dtb09}  that is about 23\% of the NE2001 distance of 1.45~kpc using  the measured  $\DM = 34.5$~pc~cm$^{-3}$.
The pulsar is located within a large ($\theta \approx 5^\circ$) extended region of elevated H$\alpha$ emission.
A clump contribution $\dDMc \sim 25^{+6}_{-14}$~pc~cm$^{-3}$ would bring  consistency with the $\tau$-DM relation. Assuming
all of  the SHASSA emission measure $\EM \sim 23~\rm pc~cm^{-6}$ is in the pulsar's foreground, the clump depth would be $\wc \gtrsim \dDMc^2 / \rm EM \sim 40~\rm pc$,  where the lower bound is the case with no internal density fluctuations.

 J1243$-$6423 (B1240$-$64) is seen through what appears to be a large
($\theta \sim 5^\circ$) region of gas   ionized by massive stars  and having an emission measure $\EM \approx 100~pc~cm^{-6}$.
The pulsar is coincident to within $\sim 1\arcsec$ of a 2MASS source
that is a stellar object in images from the SuperCOSMOS H$\alpha$ survey \citep{hmr01}.
Although pulsar timing does not indicate a pulsar companion
(R. Shannon, personal communication) the scattering deficit suggests that the DM is
enhanced by an object close enough to the pulsar to de-emphasize any scattering
contribution.

The final  three pulsars,  J1017$-$5621 (B1015$-$56),  J1932+2220 (B1930+22), and J1955+2908 (B1953+29),
are distant objects  ($d > 5~\rm kpc$) in the plane of the inner Galaxy ($|b|<0.5\deg$).  The probability is high
that the lines of sight  intersect clumps of excess free electrons or are influenced by a spiral arm local to the
pulsar.


\section{FRB Lines of Sight}
\label{sec:frbs}

 \begin{figure}[t!]
\begin{center}
\includegraphics[scale=0.45]
{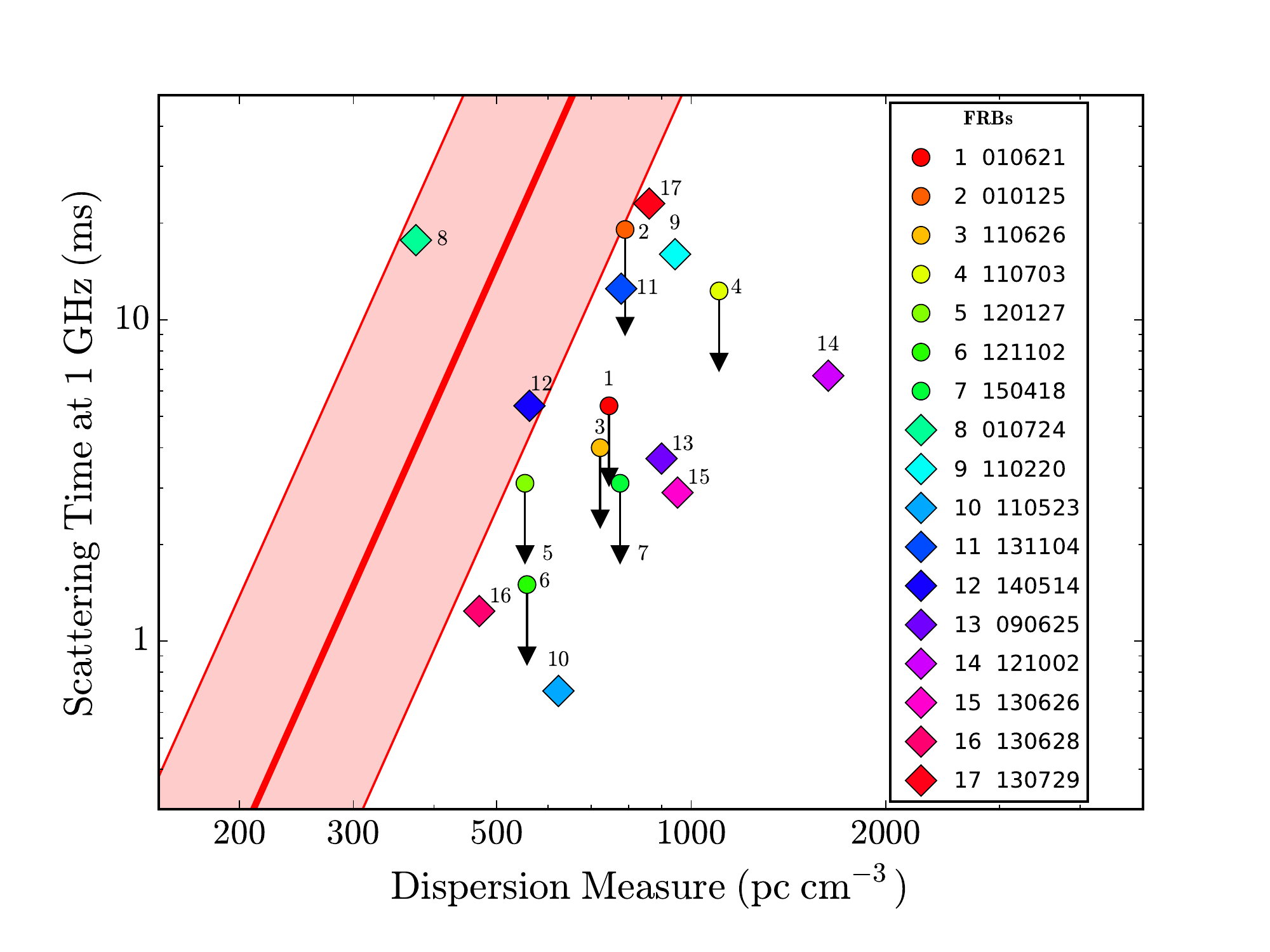}
\includegraphics[scale=0.45]
{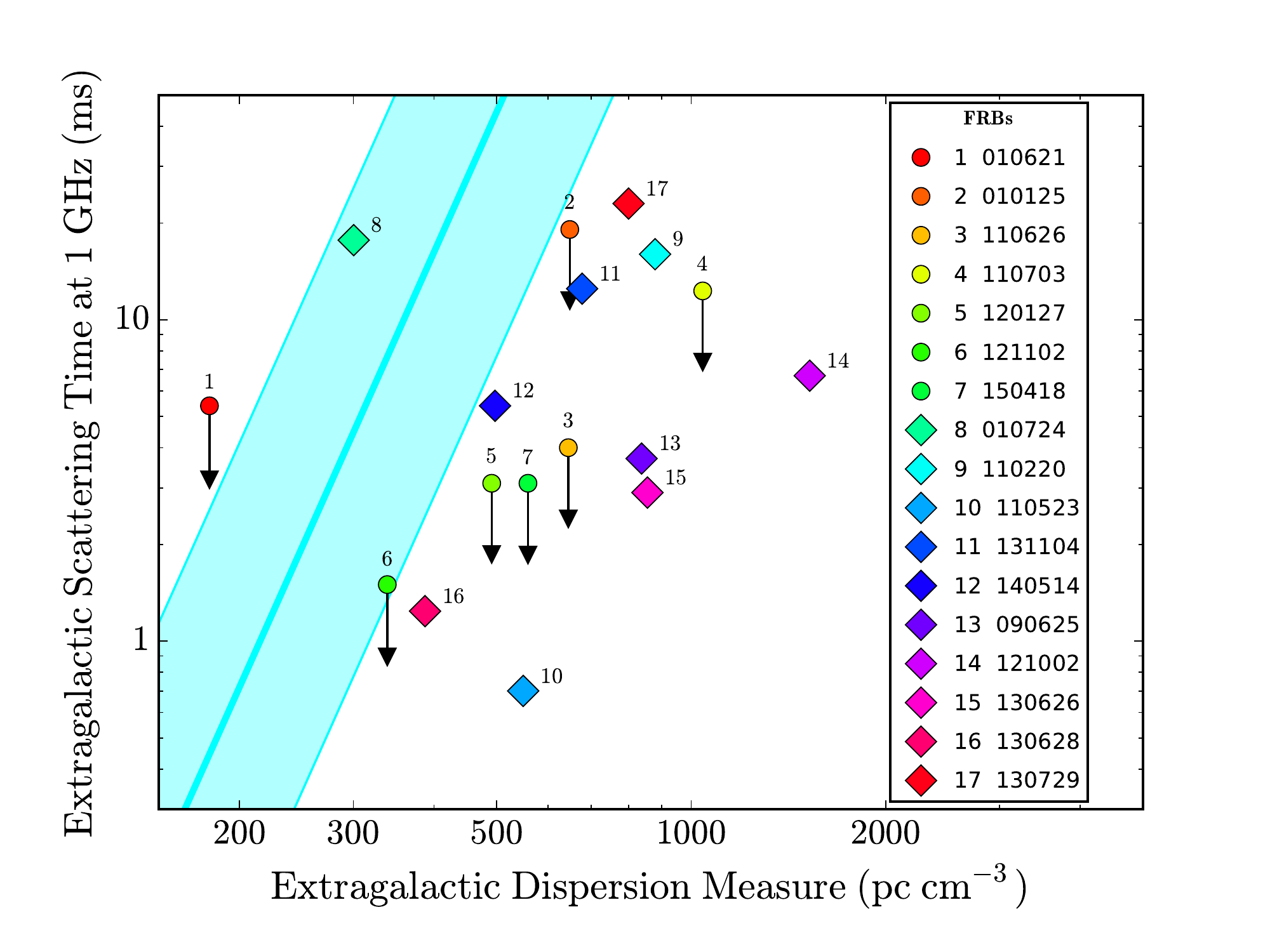}
\caption{  (Top) The $\tau$-$\DM$ plane for FRBs only using measured DM and $\tau$ values.  The red lines and shaded region show the same  model as in Figure~\ref{fig:tau-dm} and each FRB is labelled.
\label{fig:tau-dmzoom}
(Bottom) 
Similar to top panel but with the FRB values of DM and $\tau$ corrected for the contributions from the Milky Way using predictions from the NE2001 model.  Also, the  Galactic $\tau$-DM model (blue shaded region) has been shifted upward by a factor of three compared to the red region in the top panel to take into account the difference between plane and spherical waves for FRB and Galactic lines of sight, respectively.
\label{fig:taudm_mw_dm_removed_and_x3shift}
}
\end{center}
\end{figure}


The locations of FRBs in the
$\tau$-DM plane are clearly biased below the Galactic trend line.  Here we investigate the bias while  questioning whether the same trend line should apply to FRBs if their sources are extragalactic and scattering outside the Galaxy is from ionized gas with possibly different properties.

To aid our discussion,
an expanded view of the $\tau$-$\DM$ plane is shown in Figure~\ref{fig:tau-dmzoom} for FRBs only (top panel).
FRB  scattering times for different sources  vary by more than a factor of 30 but show
no obvious correlation  with their DMs. The FRBs with the largest and smallest values of $\tau$ have DMs within 25\% of each other and the largest DM has corresponding pulse broadening about equal to the median in $\log\tau$.

Assuming FRB sources are extragalactic, we define the extragalactic contribution to DM as
$\DMxgal = \DMfrb - \DMg$, where the Galactic contribution
$\DMg = \DM_{\rm NE2001}(l,b) + \DMhalo$
 is the sum of the NE2001 model integrated to its edge and a halo contribution, taken as a uniform value
$\DMhalo = 30$~pc~cm$^{-3}$ \citep[][]{2015MNRAS.451.4277D}.
Similarly we write $\tauxg = \taufrb - \taug$, where we do not include a halo contribution to $\taug$ because it is likely smaller than the
Galactic disk contribution that is itself small.
Figure~\ref{fig:tau-dmzoom} (bottom) shows FRB scattering  after the Galactic contributions to the DM and $\tau$ have  been subtracted.   In all cases, the Galactic contribution to $\tau$ is negligible so the dominant effect is a leftward shift of the points.

 In most cases the shift is small, but three objects at low Galactic latitudes  (FRB121102, FRB150418, and especially FRB010621  toward the inner Galaxy) have the  largest Galactic contributions to DM.  There are only upper bounds on $\tau$ for these three objects,  so improvements in scattering estimates, especially for the repeating FRB121102 and any others that should happen to repeat,  may move the points downward or allow actual determinations.
 
 We also note that the scattering deficit for FRBs is actually {\it larger} than it nominally appears to be in Figure~\ref{fig:tau-dmzoom} (top). 
In comparing the Galactic $\tau$-DM relation with extragalactic scattering, we need to consider differences in geometry.  For Galactic sources, which are embedded in the scattering medium, wave sphericity  causes less pulse broadening than for plane waves propagating through the same medium.   Scattered waves from a distant
source   are effectively  planar when they reach the Galaxy. This implies that scattering in the host should be a {\it factor of three  larger} for the same scattering strength, $\Ftilde \times \DM^2$.  This can be seen by comparing the ISM term in
Eq.~\ref{eq:tauism}, $\propto \Ftism\, \DMism^2 / 6$, with the host term in Eq.~\ref{eq:taufrb2}, $\propto \Fth \DMh^2 / 2$, and setting
$\DMism = \DMh$.  Scattering from a clump near the source can be as  much as a factor of six larger than for a Galactic source having the same DM.

Consider $\tauism(\DMism)$ to be the pulse broadening for a Galactic source and $\tauh$ for scattering from the distributed ISM in  a host galaxy with column density $\DMh$.
Taking the ratio of host to Galactic scattering times, we have
\be
\frac{\tauh}{\tauism} =
	\frac{3\Fth}{\Ftism}
	\left( \frac{ \DMh}{\DMism} \right)^2.
\label{eq:tauhratioh}
\ee
The analogous ratio for scattering from a clump in a host galaxy is
\be
\frac{\tauc}{\tauism} =
	\frac{6\Ftc}{\Ftism}
	\left( \frac{ \dDMc}{\DMism} \right)^2
	\left(\frac{\sc}{\wc} \right) \left(1 - \frac{\sc}{d} \right).
\label{eq:tauhratioc}
\ee

Figure~\ref{fig:taudm_mw_dm_removed_and_x3shift} (bottom)  shows the $\tau$-DM plane for FRBs with the Galactic $\tau$-DM model shifted upward by a factor of three.   In this case, only one out of the ten FRB measurements of $\tau$ are within the $\pm 1$-$\sigma$ range of
the mean Galactic model.

The leftward shifts of the plotted points  cause FRB 010724 to show {\it excess} scattering by almost one sigma and the upper limit on the low-latitude FRB 010621  exceeds the mean trend by about 1.5 sigma.    All other objects are below the mean trend line and only one
(an upper limit for FRB121102) is within one sigma. Fourteen FRBs are more than one sigma below the trend line, several by more than an order of magnitude.



\subsection{Possible Selection Effects}

A selection effect that may be relevant to the location of known FRBs in the $\tau$-DM plane  is
  the reduction in  signal-to-noise ratio for  bursts  scattered with  $\tau$ comparable to or larger than the intrinsic  pulse width (or the pulse widths that in some cases are determined by dispersion across a single spectral channel)\footnote{A reduction by 20\% occurs for 
  $\tau \approx 0.6 W$, where $W$ is the full width at half maximum of a Gaussian-shaped pulse.} This would suggest that  deficits should
occur above a horizontal line across the $\tau$-DM plane at a value of  $\tau$  larger than the typical burst width. This also  assumes that dispersion smearing across frequency channels is not a limiting factor, as with post-detection dedispersion,  which would produce a threshold that scales linearly in DM.  The distribution of DMs in the known sample   does not seem to follow either of these trends.
The same selection effect undoubtedly occurs for propagation paths through the inner disk of the Milky Way and, as previously noted,  it is likely that many sources are not seen because their bursts  propagate through long path lengths through the  host or intervening galaxies.

For the {\it known} FRB sample, however, intervening (as opposed to host) galaxies are not obviously relevant.
  A Milky Way type galaxy will scatter radiation by a minimum $\theta\sim1$~mas at 1~GHz  for a face-on geometry.  If the galaxy were midway along a $d = 1$~Gpc path, the scattering time would  be  $\tau\sim  d\theta^2/8c \sim 0.3$~s, not only much larger than observed but also  too large to allow detection of  bursts with intrinsic widths of milliseconds, unless they are extremely bright.

The agreement of some FRBs with the Galactic $\tau$-DM relation to within $\pm 1$-$\sigma$ may
 be only coincidence, but it could signify the existence of a scattering relation that is common to extragalactic and Galactic plasmas.   For these few objects there is  little or no allowance for a substantial contribution to measured DMs from electrons that do not also scatter the pulses.
This would suggest that their  scattering occurs in galaxies similar to the Milky Way from ionized gas whose internal turbulence is driven by stellar winds and supernovae.

\subsection{Dispersion and Scattering Physics for FRBs}
\label{sec:physics}


 In the Galaxy,
pulsar DMs are dominated by the warm ionized medium (WIM).  The  hot ionized medium (HIM) comprises significant volume but is a minor contributor to pulsar DMs given the approximate pressure equilibrium of the ISM, implying a  lower density for  the HIM by a factor
$T_{\rm WIM} / T_{\rm HIM} \sim 0.01$.    FRB sources may reside in regions where pressure equilibrium is not established  or where longer  path lengths encounter hot gas  that contributes significantly to DMs but has small $\Ftilde$.

The hot IGM, as many have noted, can dominate FRB DMs if their sources are at cosmological distances.
However, it  also has been argued
\citep[][]{2013ApJ...776..125M, 2014ApJ...785L..26L} that the IGM has insufficient scattering per unit DM to produce pulse broadening.
The basic problem is the very low density of the IGM.  Despite potentially large path lengths through it, a finely-tuned wavenumber spectrum
for $\delta n_e$ is needed.
Luan \& Goldreich argue that IGM heating from turbulent velocity fluctuations requires a large $\sim 1$~Mpc outer scale that prevents
density fluctuations on  small scales from being large enough to scatter radiation sufficiently.  This  corresponds to a very small
fluctuation parameter, $\Ftilde$. Even if density fluctuations are not accompanied  by turbulent velocities, thermal streaming motions in the hot ($\sim 10^6$~K) IGM would rapidly erase density variations
(1~AU / 1000~km~s$^{-1}\sim$~days).

These conclusions are not altered by  the much larger Fresnel scale
($\rF = \sqrt{\lambda \deff/2\pi}$)
 in the IGM
($\lesssim 10^{14}$~cm compared to $10^{11}$~cm for Galactic lines of sight for effective distances to the scattering medium
$\deff \lesssim 1$~Gpc and
$\deff\sim 1$~kpc, respectively).   A larger Fresnel scale  allows larger scales to diffract radiation.

The inability of the IGM to account for pulse broadening implies  that ionized regions in host galaxies must be responsible, since
intervening galaxies  are unlikely to occur along any of the known FRBs and would  produce too much scattering, as noted above.  Those regions will also contribute to DM, yielding  FRB distances that are smaller, perhaps substantially so, than estimates based solely on contributions to DM from the IGM. If scattering in  host galaxies resembles that in the Milky Way's ISM, it is justifiable to use the    Galaxy's $\tau$-DM relation as a benchmark.


We consider several effects that could influence the scattering of  FRB pulses but may be too small to produce the observed deficits in $\tau$:
\begin{enumerate}
\itemsep -1pt
\item
Pulse broadening is smaller  in the weak scattering regime, but  radio frequencies used to date   ensure that the RMS phase on the Fresnel scale
$\rF = \sqrt{\lambda \deff/2\pi}$  is larger than unity, implying that strong scattering applies \citep{r90}. Here
$\deff$ is the distance of an equivalent scattering screen, which is approximately half the FRB's distance if the IGM is involved or it can be of order the path length through a host galaxy or the location of  a subregion in a host galaxy.

\item
Dispersion and scattering that occurs  at high redshifts involves radio frequencies a factor  $1+z$ higher than the observation frequency.   This causes a host-galaxy's contribution to both DM and $\tau$ to be smaller than in the galaxy's rest frame but the scattering time depends more strongly on $z$ than DM, so a deficit may be seen.  For example, the DM from a galaxy at redshift $\zg$ is reduced by a factor $1/(1+\zg)$ but the scattering time is reduced even more by a factor  $1/(1 +\zg)^3$ \citep[][]{2013ApJ...776..125M, 2014ApJ...780L..33M, 2015MNRAS.451.4277D}. However,  since attributed redshifts are at most $\sim 1$ even if all the measured DM is attributed to the IGM,   cosmological effects  cannot by themselves account for the lack of any correlation in the $\tau$-DM plane for FRBs nor for the wide range of scattering times seen.

\item
Extragalactic plasmas may have wavenumber spectra that differ from the Galaxy's ISM.  For example,    if the  inner scale of density
fluctuations is  comparable to or larger than the Fresnel scale,  the mean square angular scattering per unit length $\eta$ is reduced by a factor $1 - (l_i / \rF)^{1/3}$
with $l_i \le \rF$ for a Kolmogorov spectrum.  Refraction from scales larger than $\rF$ can affect the arrival time of a burst but will not broaden it.     While this effect may be relevant to the IGM,  host-galaxy plasmas are probably not dissimilar from the Milky Way in their scattering properties.

\item
The lateral extent of the scattering region can limit scattering.  If it is spatially confined,  large-angle ray paths are absent and the
pulse broadening function (see Appendix~\ref{sec:app}) will not show as long an asymmetric tail as for an unlimited region
\citep[][]{2001ApJ...549..997C}.  FRBs with measured scattering  show long tails so a truncated region does not appear relevant.  However, it could apply to those FRBs that have upper limits on $\tau$.    Investigation of the detailed shapes of scattering tails may reveal or rule out this effect.

\item
Anisotropic scattering from elongated density variations alters the shape of the pulse broadening function. In the limit of a large axial ratio, the PBF is concentrated closer to the time origin.   Identification of anisotropy can be seen directly in radio images or inferred from secondary spectra of diffractive scintillations \citep[e.g.][]{2010ApJ...708..232B}.   Neither of these approaches appears feasible until fast imaging on long baselines ($>1000$~km) or single-dish measurements of dynamic spectra have sufficient sensitivity.
\end{enumerate}

%



\subsection{Dispersion and Scattering Model for FRBs}
\label{sec:frbsimple}

With the above issues in mind, we  consider FRBs in terms of  a simple model for dispersion and scattering media,
analogous to that in Figure~\ref{fig:geom_galactic} for pulsars.
Generally, the total DM of an FRB is the sum of Galactic (disk + halo), intergalactic, host galaxy, and possibly other contributions.   The host galaxy's contribution, like the Milky Way's, may also involve a mixture of smoothly distributed gas and clumps.   If any FRBs originate from high redshifts, there can also be a contribution from an intervening galaxy or galaxy cluster.

\begin{figure}[t!]
\centering
\includegraphics[trim = 30mm 35mm 15mm 35mm, clip, scale=0.575]{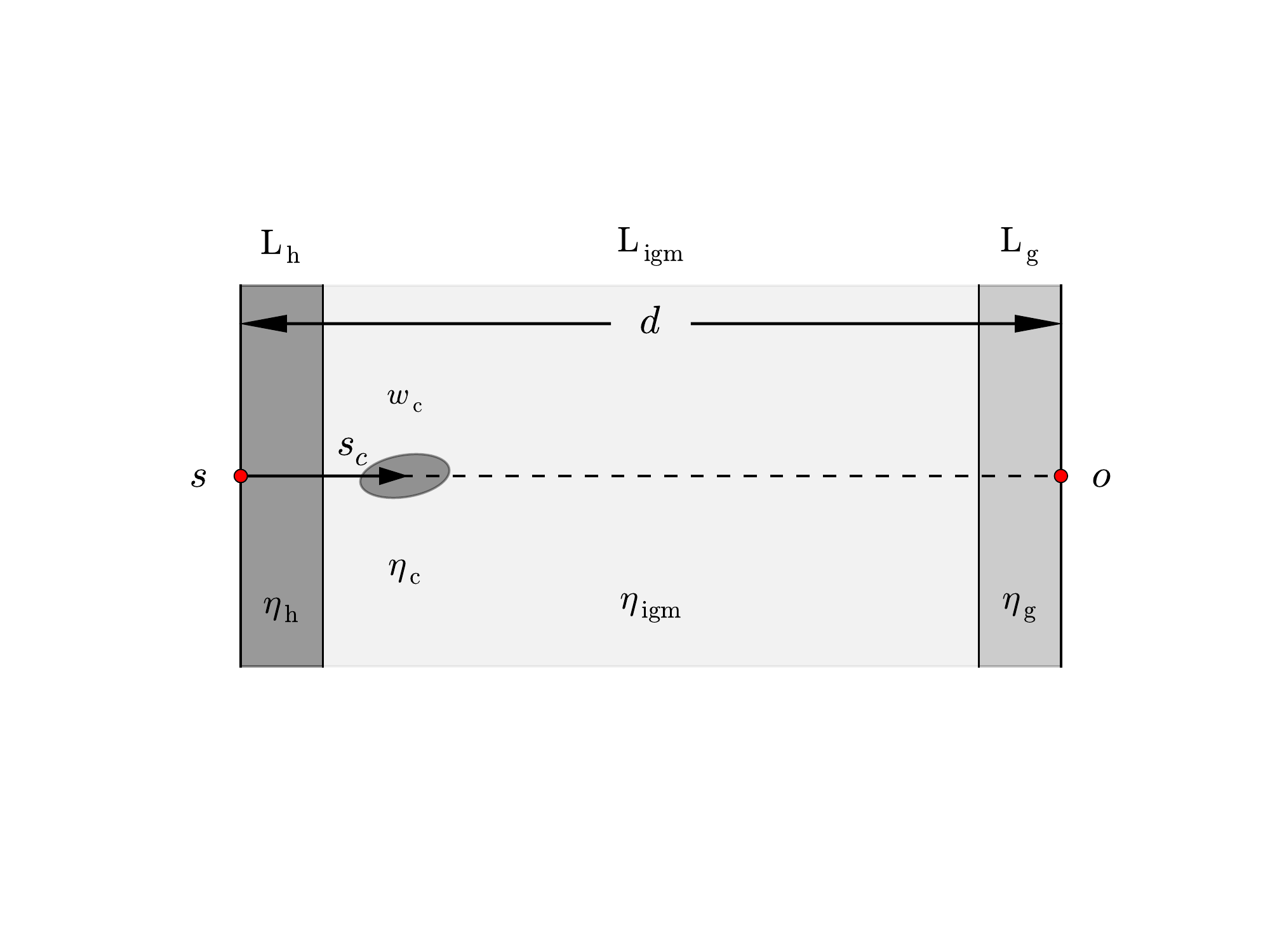}
\caption{
Schematic geometry for scattering in a host region (h), the intergalactic medium (igm), and the Galaxy (g) with thicknesses $L_{\rm h}, L_{\rm igm}$, and $L_{\rm g}$, respectively.   Extra scattering may be
contributed by a clump at distance $s_c$ from the source.
The mean-square scattering per unit distance, $\eta$ is assumed constant in each medium.
The shading corresponds to $\eta_{\rm c} \gtrsim \eta_{\rm h} \gtrsim \eta_{\rm g} \gg \eta_{\rm igm}$.
\label{fig:geom_extragalactic}
}
\end{figure}

Figure~\ref{fig:geom_extragalactic} shows the  model   we wish to consider:  a host-galaxy component,  an intervening clump region,  the IGM, and the Galaxy.   The clump component can represent substructure within the host or an intervening cloud or galaxy anywhere along the line of sight.  As with the Galactic model used in our discussion of pulsars, we assume  the local volume-averaged electron density and the mean-square scattering angle per unit length $\eta$ are constant in each component.

The total DM and pulse broadening (c.f.  Eq.~\ref{eq:2ctau}) are
\be
\DMfrb &=& \DMh + \dDMc + \DMigm + \DMg,
\label{eq:dmfrb}
\\
2c\taufrb &\approx& \etah \Lh^2/2 + \etac w_c s_c (1-s_c/d)
\nonumber \\
&&\quad\quad\quad
+ \etaigm d^2/6 + \etag\Lg^2 / 2.
\label{eq:taufrb}
\ee
The expression for $\taufrb$ includes geometric weighting for all factors and
we  have assumed that path lengths through the host ($\Lh$), clump ($\wc$), and Galaxy $\Lg$ are much less than   the path length through the IGM ($\Ligm$).   The inequalities for the different length scales are therefore $\wc \lesssim \Lg \sim \Lh \ll \Ligm \sim d$.

Using a cloud model like that  in Section~\ref{sec:psrsimple}, the scattering can be re-expressed in terms of the contributions to DM from the different components,
\be
2c\taufrb /  h_\lambda
	&=& \Fth  \DMh^2 /2
	+ \Ftc  \dDMc^2 (s_c / w_c) (1 - s_c / d)
	\nonumber \\
	&&\quad\quad+ (1/6)\Ftigm \DMigm^2
	+ \Ftg \DMg^2/2.
\label{eq:taufrb2}
\ee
The relative contributions to DM and to $\taufrb$ from the different components depend on their densities and fluctuation parameters
$\Ftilde$ and, in the case of a clump, on its proximity to the FRB source via the geometric factor $(\sc/\wc)(1-\sc/d)$.
%

\subsection{Assessment of FRB Scattering Configurations}
\label{sec:scenarios}

We now assess alternative models involving the plasma components in Figure~\ref{fig:geom_extragalactic}.
We systematically consider alternative models for FRBs with the goal of rejecting some of them and identifying tests for those that may be successful.

\subsubsection{IGM dominates $\tau$ and DM}
Suppose the IGM accounts for all extragalactic contributions to both DM and $\tau$.  Figure~4 of \citet[][]{2015MNRAS.451.4277D} shows a clear mapping between mean DM and redshift based on a simulation of  large-scale structure.    Assuming that density fluctuations scale with the local density and are driven by universal processes, a correlation between DM and scattering time is expected.  That none is seen in Figure~\ref{fig:taudm_mw_dm_removed_and_x3shift} disfavors this model on an empirical basis as well as theoretically, as described above.
A caveat is that a relatively nearby FRB population would show large scatter about any correlation due to cosmic variance.   However, a nearby population would receive a smaller contribution from the IGM to the total DM, so the caveat may not apply.
This model is also in conflict with  the observations of FRB110523 by \citet[][]{2015Natur.528..523M}, who show that the measured pulse broadening must come from within about $44~ {\rm kpc} (d / 1 \, {\rm Gpc})$ of the source  for a Gpc distance.  A source at $d \ll 1$~Gpc  would imply  an even smaller upper bound, implying that the scattering region is from substructure in a galaxy or resides in a dwarf galaxy.
\subsubsection{IGM dominates $\DMxgal$ but produces no scattering}
 A quiescent IGM   with little or no electron density variations on scales smaller than the Fresnel scale
$\rF = \sqrt{\lambda d/2\pi} \sim 10^{14}$~cm
would yield  small or negligible $\Ftigm$.   If no host galaxy or intervening clump contributes,  the measured pulse broadening times would be  produced  only by the Milky Way, and thus would be  very small for any of the known FRBs.    The  measured scattering times are much larger than the contributions from the Milky Way, so this model is also ruled out.
 \subsubsection{Quiescent IGM combined with  a host galaxy}
 Another interpretation  is like that used to account for pulsars with scattering deficits.  Here we assume that the IGM contributes to the DM but not to $\tau$.   Then scattering from a host galaxy's disk can be reconciled with the Galactic $\tau$-DM trend by simply shifting  points in  Figure~\ref{fig:taudm_mw_dm_removed_and_x3shift}  to the left by amounts that correspond to the IGM's contribution to DM (after allowance for the factor of three geometrical factor).  As before, no shift is needed for FRB010724, implying that the IGM makes a negligible contribution to its DM and that the source and host galaxy are relatively nearby.  For the other points (excluding upper limits), shifts of
$\DMigm \sim 300$ to 1000~pc~cm$^{-3}$ are needed, implying a relatively large contribution from the IGM.

 In this case
$\DMxgal = \DMigm + \DMh$.   Ignoring the low scattering from the Galaxy, the broadening time
would be  $\taudh = h_\lambda \DMh^2 \Fth / 4c$.
The question then is whether  scattering in the host galaxy follows the same $\tau$-DM trend as in the Galaxy. If so, inversion of the Galactic $\tau$-DM relation
$\taufrb = 3\taudhat(\DMh)$ yields the contribution from the host, $\DMh$, where the factor of three accounts for the difference between plane and spherical waves, as discussed earlier.
The IGM contribution is given by  $\DMigm = \DMxgal - \DMh$.    Inspection of Figure~\ref{fig:tau-dmzoom}
indicates that host contributions of 25\% to 50\% of $\DMxgal$ would give a match of the measured
$\taufrb$ to the mean Galactic trend line $\taudhat(\DMh)$.  However the large contributions from the IGM, ranging from about 50\% to 75\% of $\DMxgal$ would run afoul of the absence of low-DM FRBs if that absence persists as more FRBs are detected.
%
 %
 \subsubsection{Host galaxy  dominates both $\DMxgal$ and $\tau$}
In this case $\DMxgal = \DMh$ and scattering deficits imply that  the host galaxy does not  follow the Galactic
$\tau$-DM trend.    As discussed earlier in Section~\ref{sec:dispersion}, FRB sources distributed throughout the host galaxy would show a wide distribution of DM values, including small values well below that of the smallest
$\DMfrb = 375$~pc~cm$^{-3}$ or  $\DMxgal = 300$~pc~cm$^{-3}$.
If all of the extragalactic DM is from a host galaxy's disk  or a clump within it,  we have
$\tauh/\tauism = 3 \Ftc/\Ftism$ or $\tauc/\tauism = 6 \Ftc/\Ftism$.
The fluctuation parameter is $\Ftilde =  \zeta \epsilon^2 / \ff \left(l_i l_o^2 \right)^{1/3}$ (Section~\ref{sec:psrsimple} ).
A factor of ten scattering deficit therefore requires that the fluctuation parameter in the host be a factor of 30 or 60 smaller than in the Milky Way for a disk or clump, respectively.   In the Milky Way,  regions of more intense star formation, such as the spiral arms and  the molecular ring, have much larger  fluctuation parameters than in the ISM near the solar system.
The small fluctuation parameters inferred  here imply that the scattering regions in galaxies would be quite unusual  compared to their Galactic counterparts, corresponding to small fractional density fluctuations $\zeta \epsilon^2$  or that the product of filling factor with the  inner and outer scales of the wavenumber spectrum  is large.   The situation is most severe for FRB121002.
For those FRBs requiring $\Ftc \ll \Ftism$ or $\Fth \ll \Ftism$, it may be implausible that FRBs have dispersion measures dominated by
supernova remnants surrounding young FRB sources \citep[][]{2015arXiv150505535C}.  Galactic supernova remnants appear to increase the scattering of objects viewed through them.     What is needed in the host galaxy is a region with a combination of high electron density to provide $\DMxgal$ and high temperature to yield small density fluctuations on small scales.   Such regions would be overpressured compared to the Galaxy's ISM, suggesting an association with a galaxy center or with an ensemble of SNRs.

A counterexample is the lowest-DM FRB010724 that is  consistent with the pulsar $\tau-DM$ relation after the factor of three to six spherical-to-plane-wave effect is taken into account.   This latter object is therefore consistent with an FRB residing in a galaxy disk with scattering properties similar to that of the Milky Way.

\section{Discussion and Conclusions}
\label{sec:discussion}

This paper has compared   dispersion and scattering  of FRBs with pulsars in the Galaxy.
Pulsars show a distinct trend of increasing pulse broadening time $\tau$ with $\DM$ that we have quantified.  The scatter about the mean trend is understandable in terms of discrete scattering regions in the Galaxy and the small minority of  outliers with a deficit of scattering  can be understood in terms of scattering regions near the pulsars that produce geometrically attenuated pulse broadening.  One case appears to show an actual reduction in the turbulence within the ionized gas that dominates \DM\ \citep[][]{1992Natur.360..137P}.

FRBs as a class show deficits of pulse broadening with respect to the Galactic $\tau$-DM trend, but one FRB is consistent with the trend (FRB010724).     We discuss scattering configurations involving the Milky Way, the IGM,  host galaxies, and substructure within host galaxies that could account for the tendency for FRBs to have smaller scattering per unit DM than Galactic sources.    

Our conclusion is that pulse broadening is dominated by a host galaxy for those cases where scattering has been measured.  Upper bounds on the other FRBs allow the possibility that pulse broadening in those cases is comparable to the measured cases.  Given involvement of  a host galaxy in the scattering, 
we have identified two possible interpretations for FRBs.  
The first interpretation  associates the extragalactic portion of FRB dispersion measures  $\DMxgal$ to a mixture of IGM and host-galaxy contributions.  If scattering in the host galaxy follows the mean Galactic $\tau$-DM trend,   the host's contribution to $\DMxgal$ is
about 25\% to 50\% with the remainder coming from the IGM.   Distances or redshifts that associate all of $\DMxgal$ to the IGM will be overestimated accordingly.

The second possibility is for the host galaxy to dominate both $\DMxgal$ and $\tau$.  In this case, the electron density fluctuations
in the host must be weaker than those found in Galaxy's ISM. We quantify the scattering strength with a fluctuation parameter
$\Ftilde$ that combines the fractional variation of the density with characteristic scales of the wavenumber spectrum for the fluctuations
(Appendix~\ref{sec:measures}).   This quantity is typically a factor of 30 to 60 times smaller than in the Milky Way for lines of sight with the same DM as those of the FRBs.     To achieve these values, either the fractional density variation is much smaller than in the Galaxy or a combination of inner and outer length scales of the fluctuations is much larger.    These conditions may correspond to
 hot plasma $\gtrsim 10^6$~K that is dense enough to provide the FRB dispersion measures.

 Stronger observational constraints are clearly needed.  Localization of FRBs will yield associations with particular kinds of sources and environments and thus establish a distance scale.   A much larger sample of FRBs, along with a better understanding of selection effects in FRB surveys, will provide  DM and scattering-time distributions that can be used to disentangle the relative contributions of the IGM, host galaxies, and source environments.

\acknowledgements
We thank colleagues within the PALFA collaboration for discussions about FRBs.
We also thank Shri Kulkarni, Harish Vedantham, and Bing Zhang for organizing a meeting on FRBs held at the University of Nevada, Las Vegas in 2016 April at which some of the results of this paper were reported.
JMC thanks Michael Kramer for providing support at the Max Planck Institute for Radio Astronomy.



\begin{appendix}
\section{Scattering Data}
\label{sec:app}

Pulse broadening is measured directly on highly scattered pulsars by fitting  measured pulse shapes
with a model comprising the convolution of an  intrinsic shape with a pulse broadening function (PBF).  In the simplest scattering models, the PBF is of the form $h_d(t) \propto \exp(-t/\taud)$.   More realistic models for scattering
yield different  PBF shapes \citep[e.g.][]{lr99}.  For our analysis, alternative PBFs are not important because they yield scattering times $\taud$ that differ by no more than a factor of two, which is negligible compared to the variation of $\taud$ by ten orders of magnitude over  different lines of sight.

When  pulse broadening is too small to measure, its  reciprocal,  proportional to the scintillation bandwidth $\dnud$,  can often be obtained. Frequency structure caused by the same multipath propagation responsible for $\taud$ also produces constructive and destructive interference across the receiver bandpass \citep[][]{r90}.   Formally the scintillation bandwidth is the autocorrelation width of the frequency structure. It is related to pulse broadening through an `uncertainty' relation, $2\pi\taud\dnud = C_1$ where $C_1$ is a medium-dependent constant of order unity \citep[][]{lr99}.  We use
$C_1 = 1.16$ appropriate for a homogeneous medium with a Kolmogorov wavenumber spectrum.

Scattering times are referenced to $\nu = 1$~GHz using a $\taud\propto\nu^{-4}$ scaling law; this scaling was used in some of the data references while in other cases we applied the scaling.  It is arguable that a slightly stronger scaling  $\taud\propto \nu^{-4.4}$ should apply for low-DM objects \citep[e.g.][]{2004ApJ...605..759B} but the DM where the transition to this scaling occurs  is line of sight dependent.  Moreover, many of the original measurements were made near 1.4~GHz and, like the PBF shape, the difference in frequency scaling produces an error that   is again  small compared to the  range of scattering times  across the population.

The data used in Figure~\ref{fig:tau-dm} are part of the initial effort to aggregate scattering data on all extant data as input to the next generation Galactic electron-density model that will supersede
the NE2001 model \citep[][]{2002astro.ph..7156C, 2003astro.ph..1598C}.  Here we describe references to the data.

Many of the data points  were used in the development of the NE2001 electron density model
and included direct measurements of pulse broadening $\taud$ and indirect measurements calculated from   diffractive interstellar scintillation (DISS)  bandwidths $\dnud$  using  $\taud = C_1 / 2\pi\dnud$.   In the figure the two kinds of points are separately designated.
The NE2001 papers contain bibliographies of primary source data.

Pulse-broadening measurements and upper limits were also used from sizable samples
reported by
\citet{2004ApJ...605..759B},
\citet{2013ApJ...772...50N},
\citet{kmn15},
and for the individual pulsars
J1811$-$1736
\citep{2007A&A...462..703C},
J1841$-$0500
\citep{2012ApJ...746...63C},
J2021+3651
\citep{2004ApJ...612..389H},
and
J2022+3842
\citep{2011ApJ...739...39A}.
Of these, J1841$-$0500 (DM = 532~pc~cm$^{-3}$) has one of the two  largest known pulse broadening times, $\taud = 2.3$~s at 1 GHz, the object being one of the GC pulsars,  J1745$-$2912.
 We have used some of the pulse broadening times from \citet[][]{2015MNRAS.454.2517L} that includes results from
\citet[][]{2013MNRAS.434...69L} and \citep[][]{2015ApJ...808...18L}.  There is considerable overlap with the NE2001 sample, which also already included pulsars observed by  \citep[][]{2001ApJ...562L.157L, 2004A&A...425..569L}. The pulsar J1740+1000 is also included \citep[][]{2002ApJ...564..333M} as an estimate of pulse broadening from a DISS bandwidth measurement.  The value used is a factor of more than $10^4$ smaller than the pulse broadening time reported by \citet[][]{2013MNRAS.434...69L}, which appears to be confused with asymmetry from intrinsic pulse structure, as caveated by  \citet[][]{2013MNRAS.434...69L}.

Additional scintillation bandwidths were used from \citet{2013MNRAS.434...69L},
\citet{2013MNRAS.429.2161K}, and \citet{2016ApJ...818..166L}.
The lone LMC pulsar with a scattering measurement is J0540$-$6919 (B0540-69) \citep[][]{2003ApJ...590L..95J}.

For the five pulsars in the Galactic center, we use reported estimates or derive our own upper limits from  pulse profiles reported in
\citet{2006MNRAS.373L...6J}
and
\citet{2009ApJ...702L.177D}.
For the magnetar J1745$-$2900 near the Galactic center, we use the pulse broadening time reported in
\citet{2014ApJ...780L...3S}.

For FRBs we used  individual references for the seventeen objects available in the literature as of 2016 March 1:
\citet[][]{
2007Sci...318..777L,			
2012MNRAS.425L..71K, 		
2013Sci...341...53T,			
2014ApJ...789L..26P,		
2014ApJ...790..101S,		
2015MNRAS.447..246P,		
2015ApJ...799L...5R,		
2015Natur.528..523M,		
2016Natur.530..453K,		
2015arXiv151107746C}		
and we referred to the online FRB catalog
(Petroff et al., in preparation;
\url{http://www.astronomy.swin.edu.au/pulsar/frbcat/}.

\section{Clump Dispersion, Scattering and Emission Measures}
\label{sec:measures}

We give a short summary of the relationships between the dispersion measure, scattering measure, and emission measure. For more details see \citet[][]{1991Natur.354..121C, 1993ApJ...411..674T, 2002astro.ph..7156C}.

Consider a clump with internal, volume-averaged electron density $\nec$.
If the clump comprises  sub-clumps with internal density $\nesc$ and filling factor $\ff$, then $\nec = \ff\nesc$.
Within each sub-clump  there are small-scale density fluctuations  consistent with a power-law  wavenumber spectrum  $\cnsq q^{-\beta}$
 having a smallest (inner) scale $l_i= 2\pi/q_i$ and largest (outer) scale $l_o= 2\pi/q_o$.
 In the paper we use $\beta = 11/3$ for a Kolmogorov wavenumber spectrum.
 Integrated over all wavenumbers $q$ to get the variance, the fractional density fluctuation
is $\epsilon = $~(RMS density) / (mean density).
Variations between sub-clumps are described by the dimensionless second moment
$\zeta = \langle \nesc^2 \rangle / \langle \nesc \rangle^2$, where angular brackets denote
averages over an ensemble.

For a path length $\wc$ through the clump , the dispersion measure
 and scattering measures are
\be
\DMc &=& \nec \wc
\\
\SMc &=& \int_{\rm clump} \!\!\!\!\!\!\!\!\!\! ds\, {\cnsq}(s) =
	\CSM
	\left( \frac{\Fc\DMc^2}{\wc} \right),
\ee
where $\CSM =  \left[3(2\pi)^{1/3}\right]^{-1}$ and
the `fluctuation' parameter is
$\Fc = \zeta\epsilon^2 / \ff l_{\rm o}^{2/3}$.

The emission measure is the integral through the clump of the square of the electron density,
$\EMc = \int ds\, n_{\rm e}^2(s)$,
that includes all fluctuations.   It can be related to $\DMc$ using
\be
\EMc &=& \frac{\zeta \left( 1 + \epsilon^2 \right) \nec^2 \wc}{\ff}
= \frac{\zeta \left( 1 + \epsilon^2 \right) \DMc \nec }{\ff}
= \frac{\zeta \left( 1 + \epsilon^2 \right) \DMc^2 }{\ff\wc}.
\label{eq:EMc1}
\ee
It can also be related to a combination of $\DMc$ and $\SMc$,
\be
\EMc =
\left(\frac{\zeta}{\ff} \right) \frac{ \DMc^2 }{\wc} + \frac{\displaystyle l_{\rm o}^{2/3} \SMc}{\CSM}
=
\left(\frac{\zeta}{\ff}\right)  \frac{ \DMc^2 }{\wc} +  544.6\ {\rm pc\ cm^{-6}} \ l_{\rm o}^{2/3}(\rm pc) \, \SMc ,
\label{eq:EMc2}
\ee
where the second equality gives EM in standard units of pc~cm$^{-6}$ when $\wc$ and $l_{\rm o}$ are both in pc and
$\DMc$ and $\SMc$ are expressed in their  standard units of pc~cm$^{-3}$ and kpc~m$^{-20/3}$, respectively.
Equations~\ref{eq:EMc1} and \ref{eq:EMc2} imply that density fluctuations internal to the clump can increase $\EMc$ substantially.   For 100\% fluctuations ($\epsilon = 1$),  $\EMc$ is at least double the value that results from using  the expression $\EMc = {\DMc}^2 / {\wc}$, as is commonly assumed in the literature and which is only a lower bound on $\EMc$. Cloud-to-cloud variations quantified by $\zeta\ge 1$ with a non-unity filling factor $\ff\le1$ can make $\EMc$ even larger.  The $\SMc$ term accounts for Kolmogorov fluctuations in $\nec$.

We evaluate the emission measure for a region of depth $L$, 
\be
{\EM} = 10^3\,{\rm pc\ cm^{-3}}\, \frac{\zeta(1+\epsilon^2)}{{\ff}} \left( \frac{\DM_{1000}^2} {L_{\rm kpc}} \right),
\ee
implying a a free-free optical depth for temperature $T_{e} = 10^4T_{\rm e, 4}$~K,
\be
\tau_{\rm ff} = 3.3\times10^{-3} \nu^{-2.1} T_{\rm e,4}^{-1.35} \frac{\zeta(1+\epsilon^2)}{\ff} \left( \frac{\DM_{1000}^2} {L_{\rm kpc}} \right).
\ee

A clump with a broad spectrum of density fluctuations will both diffract and refract incident radiation.
Diffraction causes radiation to be scattered instantaneously into a range of directions.   Refraction
simply bends a ray path and does not broaden an image or give a spread in arrival times, though it will produce a delay in arrival time of an incident pulse.    The Fresnel scale
$\rF = \sqrt{\lambda  d_{\rm se}/ 2\pi}$ separates diffraction from refraction if plane waves are incident on a single electron-density screen at distance $d_{\rm se}$ from the observer.  However a source at finite distance $d$ and a screen at distance $d_s$ from the source has a Fresnel scale
$\rF = \sqrt{\lambda d/2\pi} \left[\left(d/\ds\right) \left(1-\ds/d\right)\right]^{1/2}$.

In the main text we  use  the mean-square scattering angle per unit length $\eta$, which is
\be
\eta =  \frac{\lambda^4 r_e^2 \Gamma(3-\beta/2)} {4-\beta} q_i^{4-\beta} \cnsq
\ee
for  $l_i \ll \rF$  \citep[][]{1998ApJ...507..846C}.
For $\beta = 11/3$ and defining
$h_\lambda = \Gamma(7/6) \lambda^4 r_e^2$ , we have
\be
\eta = h_\lambda \Ftilde_c\nec^2,
\ee
where $\Ftilde_c = \Fc / l_o^{2/3}$ is the modified fluctuation parameter used in the main text.

If the inner scale $l_i$ is not much smaller than the Fresnel scale,
$\eta$ is reduced by a factor $\left[1 - (l_i / \rF)^{4-\beta} \right]$  for $l_i \le \rF$,
\be
\eta =  h_\lambda \Ftilde_c\nec^2 \left[1 - (l_i / \rF)^{4-\beta} \right].
\ee
This result implies that diffraction, which causes pulse broadening, diminishes as the inner scale
approaches the Fresnel scale.    Refraction from larger scales still occurs, but it causes time-of-arrival perturbations on long time scales rather than causing multipath propagation that instantaneously broadens a pulse.    It should be noted that refraction much stronger than from a Kolmogorov medium can produce multiple images that would yield multiple pulses with different arrival times and thus mimic pulse broadening from diffraction.

\end{appendix}
\end{document}